\documentclass[12pt,preprint]{aastex}

\shorttitle{Catalog of Galaxy Morphology in Clusters}
\shortauthors{Saintonge et al.}

\begin{document}

\title{Catalog of Galaxy Morphology in Four Rich Clusters : Luminosity Evolution of Disk Galaxies at $0.33<z<0.83$\footnote{Based on observations 
obtained with the NASA/ESA \textit{Hubble Space Telescope}, which is operated by STScI for the 
Association of Universities for Research in Astronomy, Inc., under NASA contract NAS5-26555.}}

\author{Am\'{e}lie Saintonge\footnote{Current address: Center for Radiophysics and Space Research, Cornell University, Space Sciences Building, Ithaca, NY 14853; {\it amelie@astro.cornell.edu}},  David Schade}
\affil{Dominion Astrophysical Observatory, Herzberg Institute of Astrophysics, National Research Council of Canada, 
Victoria, BC, V9E 2E7, Canada}

\author{E. Ellingson}
\affil{Center for Astrophysics and Space Astronomy, University of Colorado, Boulder, CO, 80309}

\author{H.K.C. Yee and R. G. Carlberg}
\affil{Department of Astronomy, University of Toronto, Toronto, ON, M5S 3H8, Canada}

\begin{abstract}

\textit{Hubble Space Telescope} (HST) imaging of four rich, X-ray luminous, galaxy clusters ($0.33<z<0.83$) is
used to produce quantitative morphological measurements for galaxies in their fields. 
Catalogs of these measurements are presented for 1642
galaxies brighter than F814W(AB)=23.0 .
 Galaxy luminosity profiles are fitted with three models: exponential disk, de Vaucouleurs bulge, and a
disk-plus-bulge hybrid model.  The best fit is selected and produces a quantitative
assessment of the morphology of each galaxy: the principal parameters derived being $B/T$, the ratio
of bulge to
total luminosity, the scale lengths and half-light radii, axial ratios,
position angles and surface brightnesses of
each component. Cluster membership is determined using a statistical correction for
field galaxy contamination,
and a mass normalization factor (mass within boundaries of the observed fields) 
is derived for each cluster.
Morphological classes are defined using $B/T$: disk galaxies have $0\leqslant B/T \leqslant0.4$,
intermediate galaxies $0.4 < B/T < 0.8$, and bulge-dominated galaxies have $0.8\leqslant B/T \leqslant1$.
 In the present paper, this catalog of measurements is used to investigate the luminosity evolution of disk
galaxies in the rich-cluster environment. Examination of the relations between disk scale-length
and central surface brightness suggests, under the assumption that these clusters represent a family
who share a common evolutionary history and are simply observed at different ages, that there
is a dramatic change in the properties of the small disks ($h < 2$ kpc). This change is
best characterized as a change in surface brightness by $\sim 1.5$ magnitude between $z=0.3$ and $z=0.8$
with brighter disks at higher redshifts. 

\end{abstract}

\keywords{galaxies: evolution, galaxies: spiral, 
galaxies: clusters: individual (MS 1358.4+6245, MS0015.9+1609, MS 1621.5+2640, MS 1054.4-0321)}

\section{INTRODUCTION}

Because of their extreme environments, clusters are interesting places in which to study galaxy evolution \citep[]{dress84,martel98}.  
Their cores have the highest volume
density of galaxies in the Universe so that any environmental dependence of galaxy formation or
evolution processes should be most pronounced there, when contrasted with
studies of the field population. A more
pedestrian motivation for studying galaxies in clusters is that the high surface density of objects makes them
easy targets for imaging and multi-object spectroscopy. Reasonably well-selected samples of clusters are
now available up to redshift approaching $z=1$ (e.g. \citet[]{rosati98,gladders02}). 
However, it could be
argued that the sample of well-studied clusters (those with high-resolution imaging, morphological measurements,
 and extensive spectroscopy)
is still fairly small. The small sample means that, if galaxy clusters are a diverse
population (in terms of richness, X-ray luminosity, behavior of galaxy populations), then it will be difficult
to draw the correct general conclusions from studying the present sample. 
Furthermore, it is difficult to trace the pedigree of
the sample of clusters that has been well-studied to this date. By this we mean that a complex and tangled
process of selection has been applied to broader samples of clusters and that process results in the
well-studied sample that presently exists. If, for example, a cluster was chosen for extensive
spectroscopy because of the presence of a large fraction of blue galaxies, and then chosen for HST imaging
because it had a large population of emission-line galaxies, then this cluster cannot be claimed to be
a member of a representative set of clusters. Conclusions drawn from a sample with this type of pre-selection
for detailed study will not be generally applicable.
 The CNOC \citep[]{CNOC1} sample of clusters was chosen by X-ray luminosity and redshift and should avoid
some of these potential problems.

The phenomenology of galaxy populations in clusters can be divided roughly (and perhaps not physically)
into five areas. The first is the formation and evolution of elliptical galaxies which dominate the
core of clusters \citep[]{hubble31} and which may have been in place prior to cluster virialization
\citep[]{dress97}. Ellipticals seem to form the backbone of the cluster galaxy population. In local clusters
they show a very tight color-magnitude relation with small scatter which
implies either an early formation epoch or synchronization of formation times \citep[]{bower92}.
Nearby ellipticals follow a tight fundamental plane relation between size, surface brightness, and velocity
dispersion \citep[]{djorg87,dress87}.  The tightness of the cluster
color-magnitude relation seems to be preserved \citep[]{ellis97} to $z\sim 0.5$ placing tighter constraints
on the formation epoch. Studies of the moderate redshift fundamental plane \citep[]{vanD98b,kelson} and the
relation between size and luminosity \citep[]{Schade97}, a projection of the fundamental plane, indicate
that cluster ellipticals are evolving passively to $z\sim 1$, that is, their stellar populations are aging with
little ongoing star formation. The data that are available at the present time indicate that all
ellipticals in all clusters that have been studied evolve in a similar fashion. Interestingly,
cluster and field elliptical populations seem to evolve in an identical manner \citep[]{Schade99}.

The second observed cluster
phenomenon is the morphology-density relation
or morphology-clustercentric radius
relation \citep[]{mel77,dress80,whit93}.
These relations describe how the relative fraction of different morphological types varies rapidly
with distance from the cluster core or with the local galaxy density. Cluster
cores are dominated by elliptical and S0 galaxies whereas the outer, lower-density regions more nearly
approach a spiral-rich mix of types similar to the field. It is still debated whether the
dependence on density or distance from the cluster center is more fundamental. Dressler (1997) compared
the morphology-density relation in local clusters with those at moderate redshift. When clusters of
all types are viewed together the morphology-density relation is nearly absent at $z=0.5$ except for
the regions of highest density where ellipticals dominate. In contrast, the low redshift sample
of all cluster types shows a clear and continuous morphology-density relation. If, however, the clusters
are divided into samples of high-concentration regular clusters and low-concentration or
irregular clusters the situation is different. At low redshift the morphology-density relation is
observed in clusters of all types whereas, at $z=0.5$, only the high-concentration clusters show
a clear relation. Irregular clusters show no dependence of the fractions of any particular
morphological type on local density.

The
Butcher-Oemler (B-O) effect \citep[]{bo84} is the third phenomenon related to cluster galaxy populations
but the first to be reported.  \citet[]{bo84} studied a sample of 33 clusters with
$0.003 < z < 0.54$ and found a rapid increase in the fraction of blue galaxies (defined
as those with B-V colors more than 0.20 mag bluer than the ridge line of early-type galaxies)
with redshift. In the local Universe, the fraction of cluster galaxies that meet this
criteria is a few percent whereas, at $z=0.5$, some clusters have blue fractions approaching
35\% (although the redshift dependence is not uniform for all clusters). The effect continues
to $z\sim1$ \citep[]{rakos95} but there is a wide range of values even at
low redshift.
 \citet[]{lavery92} and \citet[]{lavery94} found,
from high-resolution ground-based imaging, that  many of the Butcher-Oemler
galaxies were disk galaxies with widely distributed star formation (as opposed to
concentration toward the nucleus) and that galaxy-galaxy interactions appear to be
responsible for the enhanced star formation in some of the systems. These effects were
confirmed by HST imaging \citep{dress94,couch94}. \citet[]{oemler97}
called into question a direct link between blue color and interactions and pointed out
the possible effect of magnitude-selection on the spectroscopic samples.

 The fourth problem of phenomenology is the role of interactions, mergers, or
``galaxy harassment'' \citep[]{moore96} in the evolution of cluster galaxy population. There are actually
two problems. The first is the effect of interactions between galaxies and the intra-cluster
medium (ICM) and the second is the effect of galaxy-galaxy interactions on the evolving population.
The velocity dispersion in rich clusters is large ($\sim 1000$ km/sec) so that galaxy-galaxy
encounter velocities are typically too high for actual merging to be a probable outcome of galaxy
encounters. Still there exists the
possibility of modifying galaxy morphology by close encounters.
Galaxy ``harassment'' \citep[]{moore98} has been proposed as a potentially important driver
of cluster galaxy evolution. Many studies note the
large fraction of apparently interacting galaxies, for example \citet{lavery92,dress94} and \citet{couch94}.
But the absence of a correlation between nearest neighbor distance
and color (indicative of star formation activity) in clusters  \citep{oemler97} is puzzling.
In contrast, \citet[]{vanD99}
claim direct evidence in a cluster at $z=0.83$ for merger-driven production of massive
early-type galaxies. The merger product progenitors are typically red, early-type galaxies so
that they do not provide part of the solution to the B-O problem.

The most
recently defined (and fifth) phenomenon to come into focus is the so-called ``S0 problem''.
In the study of the morphology-density relation by \citet[]{dress97}
 it was found that although the fractions of galaxies of type S0 in clusters show little dependence
on galaxy density either locally or at $z=0.5$ (regardless of cluster type), the overall
fraction of the galaxy population contained in S0s is much lower ($\sim 20\%$) at moderate redshift
than in clusters in the local Universe (where S0s constitute $\sim 45\%$ of the population). The spiral
fraction varies in such a way to roughly balance this change with redshift.
These two facts taken together suggest that some process is causing the transformation from spirals
into S0s from $z\sim0.5$ to $z\sim0$.

 Our understanding of star formation in clusters---which clearly has important ramifications for
all of the observational phenomena described above---has improved, to some degree, in recent years.
It has been shown clearly \citep[]{balogh97} that star formation is suppressed in clusters relative
to the field and that this effect is present over the entire cluster volume out to $\sim$ twice the
virial radius. Furthermore, this effect is not due to the existence of the
morphology-density relation. A comparison of cluster and field galaxies with matched sizes, and
ratios of bulge-to-total luminosities \citep[]{balogh98} shows that the cluster galaxies have distributions with
lower mean star-formation rates. Large analysis of spectroscopic data to study the frequency of
star-forming and post-starburst galaxies have been done by \citet[]{balogh99}  and by \citet[]{poggianti99}.  
There is real disagreement in these studies about whether post-starburst galaxies
are more common in clusters relative to the field which would indicate the truncation of
star-formation upon infall into the cluster.  \citet[]{balogh00} present a modeling
of the ongoing accretion of field galaxies with their star-formation declining on a
gas-consumption timescale after their gas reservoirs have been stripped off by interaction
with the ICM. The curious observation that the suppression of star formation occurs out
to large distances from the cluster core is explained by the feature of their N-body simulations
that many galaxies observed as far out as twice the virial radius have, in fact, visited the
central regions of the cluster in the past.

 The present study presents a large catalog of photometric and morphological measurements of a set of
four X-ray luminous clusters spanning a wide range in redshift. A preliminary analysis of one of the aspects
of galaxy evolution in clusters is presented.

A strong emphasis is put here on describing the fitting and analysis techniques employed, but an examination of
the luminosity evolution of small disk galaxies is also presented.  
In \S \ref{data} a description of the data selected for this study and
their reduction is made, and the method used to get a quantitative description of the galaxy morphology is presented
in \S \ref{morphology}.  Finally, in \S \ref{results} and \S \ref{discussion} are presented and discussed the results of the study.  The discussion is centered
around the main result: luminosity evolution.  However, other questions are raised, such as the color-magnitude
and morphology-density relations, to verify the validity of the fitting technique and the reliability of the
classification method applied on the galaxy sample.

 All cosmology-dependent results in this paper are derived using $H_\circ=70$ km sec$^{-1}$ Mpc$^{-1}$, 
$\Omega_m=0.3$ and $\Omega_{\Lambda}=0.7$.

\section{DATA \label{data}}

\subsection{Cluster Selection}
The CNOC (Canadian Network for Observational Cosmology) cluster redshift survey
\citep[]{CNOC1} is a study of 16 galaxy clusters with X-ray luminosity in excess of 
$4 \times 10^{44}$ ergs s$^{-1}$ and with redshifts covering the
range $z=0.2$ to $z=0.55$. Such clusters are likely to be rich and virialized.
 The survey consists of imaging data and redshifts for approximately
2600 faint galaxies. A number of important results have been derived from this dataset, ranging
from a determination of $\Omega_{m}$ and  $\sigma_{8}$ \citep[]{carlberg97a,carlberg1}
to studies of evolution of galaxies (e.g. \citet[]{Schade96cnocE,ellingson01}).

Three clusters from the CNOC sample were chosen for further imaging with HST.  They were 
selected out of the 16
in order to cover an interesting redshift range.  They are MS1358.4+6245 \citep[]{CNOC4}, MS1621.5+2640 
\citep[]{CNOC3}, 
and MS0015.9+1609 \citep[]{CNOC6} (hereafter MS1358, MS1621, and MS0016, respectively). Existing
archival imaging was supplemented with new imaging with the intent of sampling galaxies at a variety
of distances from the cluster center.  The cluster redshifts  
are presented in Table \ref{tab1}.  In order to cover a wide range of
redshifts, a fourth cluster is added to the present study, cluster MS 1054.4-0321 (MS1054) 
at $z=0.832$ \citep[]{vanD99}.
MS1054 is also a X-ray cluster, and so has been selected on the same basis as the three
CNOC clusters.  MS1054 and the three CNOC clusters are part of the EMSS (\textit{Einstein Observatory}
Extended Medium-Sensitivity Survey, \citet[]{emss}), and their X-ray fluxes are presented in Table \ref{tab1} (note, however, that a recent reanalysis of the X-ray luminosities of a cluster sample with $0.3<z<0.6$ including our three CNOC clusters, suggests that the values in the EMSS are underestimated \citep[]{ellis02}).

 The cluster galaxy sample in the present paper has an excellent redshift baseline ($z=0.3$ to $z=0.8$)
and thus is suitable for the study of the change in galaxy properties with redshift.
 Secondly, it samples these clusters over a range of distance from the cluster center so that
the effect of galaxy environment can be taken into account. 

\subsection{Imaging and Reduction}
The original imaging of the CNOC clusters was obtained at CFHT using the multi-object spectrograph
and this imaging was of moderate quality with seeing largely in the range 0.9 to 1.1 arcseconds and 
significant variations in the point-spread function within the frames.
For the present 
study, additional data was acquired
with the \textit{Hubble Space Telescope} for the three selected clusters.
Clusters MS0016, MS1358, and
MS1621 were observed with the WFPC2
and each field was observed with two filters: F814W and either F450W or F555W. 
 To this data core was added previous WFPC2 observations of these
clusters and of MS1054.  A large quantity of data \citep[]{kelson} was produced for MS1358 
 in the F814W and F606W bands.  The same filters were used to observe
MS1054 \citep[]{vanD99}.  Some additional exposures
or MS0016 were taken with F814W and F555W (Proposals 8020 and 5378)
\footnote{The data were retrieved through the Canadian Astronomy Data Centre, which is operated 
by the Herzberg Institute of Astrophysics, National Research Council of Canada}.
Figure \ref{radec} shows the distribution of the galaxies in each cluster, resulting from the
combination of all these data sets.  The circles represent the characteristic radius of each cluster, 
$r_{200}$, the distance from the center where the average density is two hundred times the 
critical density of the Universe (the values of $r_{200}$ are given in Table \ref{tab1}).

The data were reduced by an automated pipeline, developed at the Canadian Astronomy Data Centre, that 
takes as input an ``association'' name (defined as a group of images that can be co-added) and determines
the offsets between the images (the ``dither'' pattern) and the other information needed to execute
the stacking and sky subtraction. This pipeline is composed largely of components from a number
of STSDAS 
\footnote{STSDAS is a product of the Space Telescope Science Institute, which is operated by AURA for NASA} and IRAF \citep[]{iraf}  packages.  
For the present study images were simply shifted and averaged.
The performance of this pipeline was verified by repeating the processing for the data in this paper
independently in a manual mode and inspecting the results at each stage. The output of the pipeline
is the stacked and cosmic ray-rejected image. Corrections for systematic errors in the astrometry
derived from the HST images using their WCS information are also made automatically by this pipeline.

A number of minor image anomalies had to be corrected before the data were analyzed.  
The cluster MS1358+62 was observed in the continuous viewing zone of HST
\citep[]{vanD98} , and most images showed large shadow stripes along
the diagonals of the frame.  To correct for this effect,
 the frames were rotated by the
appropriate angle to position the stripes horizontally or vertically.  Then
using a long and thin median filter box, filtered images representing the
shadow pattern were obtained and then subtracted from the original images.
The filtered images were produced by the MEDIAN task in IRAF, which acts by
replacing the central pixel of the box
by the median of all the pixels in that window.  The shape of the box ($500 \times
5$ pixels) allowed the median to be computed only on the sky, leaving out the
objects.  After applying this correction, the sky value is more uniform, and
no significant noise enhancement is noted in the previously shadowed regions.
Most of the exposures also had to be corrected for bad pixels. 

Photometric zeropoints in the AB system were calculated using the standard PHOTPLAM and
PHOTFLAM parameters of HST imaging.  The values taken are from the last updated
tables\footnote{The tables can be found at ${\tt http://www.stsci.edu/instruments/wfpc2/Wfpc2\_phot/wfpc2\_photlam.html}$} , 
and the zeropoints are obtained by:
\begin{displaymath}
zp=-2.5\log{\left( \begin{array}{c} {\tt photflam} \end{array} \right)}-2.5\log{\left( \begin{array}{c} {\tt photplam}^{2}/3\times10^{8} \end{array} \right)}-48.60
\end{displaymath}
These magnitude zeropoints were then corrected to account for Galactic extinction
with factors produced by the NASA/IPAC Extragalactic Database (NED) (values are presented in Table \ref{tab1} for $V$-band extinction, under $A_{V}$).

\subsection{Galaxy Selection}

A catalog of galaxies was created using the SExtractor software \citep[]{bert96}.  
Before doing so, the borders of the frames along the edges with insufficient signal were masked and
were filled with random noise with the same dispersion as that of the actual image.  This was done 
to avoid spurious detections that are produced by SExtractor
along the edges where there is an abrupt transition from image noise to zero noise
in the borders.  The noise was generated using the MKNOISE task in IRAF. This procedure yields
a cleaner, and no less complete, catalog of galaxies.

 In order to determine the probability of detection as a function of galaxy type, 
simulations of the detection process were carried out. A range of galaxy types,
sizes, and central surface brightness were created, convolved with the observed point-spread
function, and distributed at random across the actual frames. Crowding leads to a
loss of detection sensitivity and the number of galaxies simulated on each frame was
limited to a reasonable number (5-10 per image) to avoid unrealistic levels
of crowding. Several thousand galaxies of each type with a range
of brightness were simulated for each of the clusters.

 The signal-to-noise ratio at a given apparent magnitude varies from cluster to
cluster and, in some cases, from image to image within a cluster. Simulations were
made over all actual images of the clusters. SExtractor was run on the images that
contained all of the real galaxies plus the added simulated galaxies and the list
of galaxies detected by SExtractor was compared to the input list of simulated 
galaxies. The detection probability for a given galaxy type, size, and brightness is
the ratio of number of galaxies detected to the number of galaxies inserted into
the images. This number is averaged over all of the fields of a given cluster.

\begin{figure}
\plotone{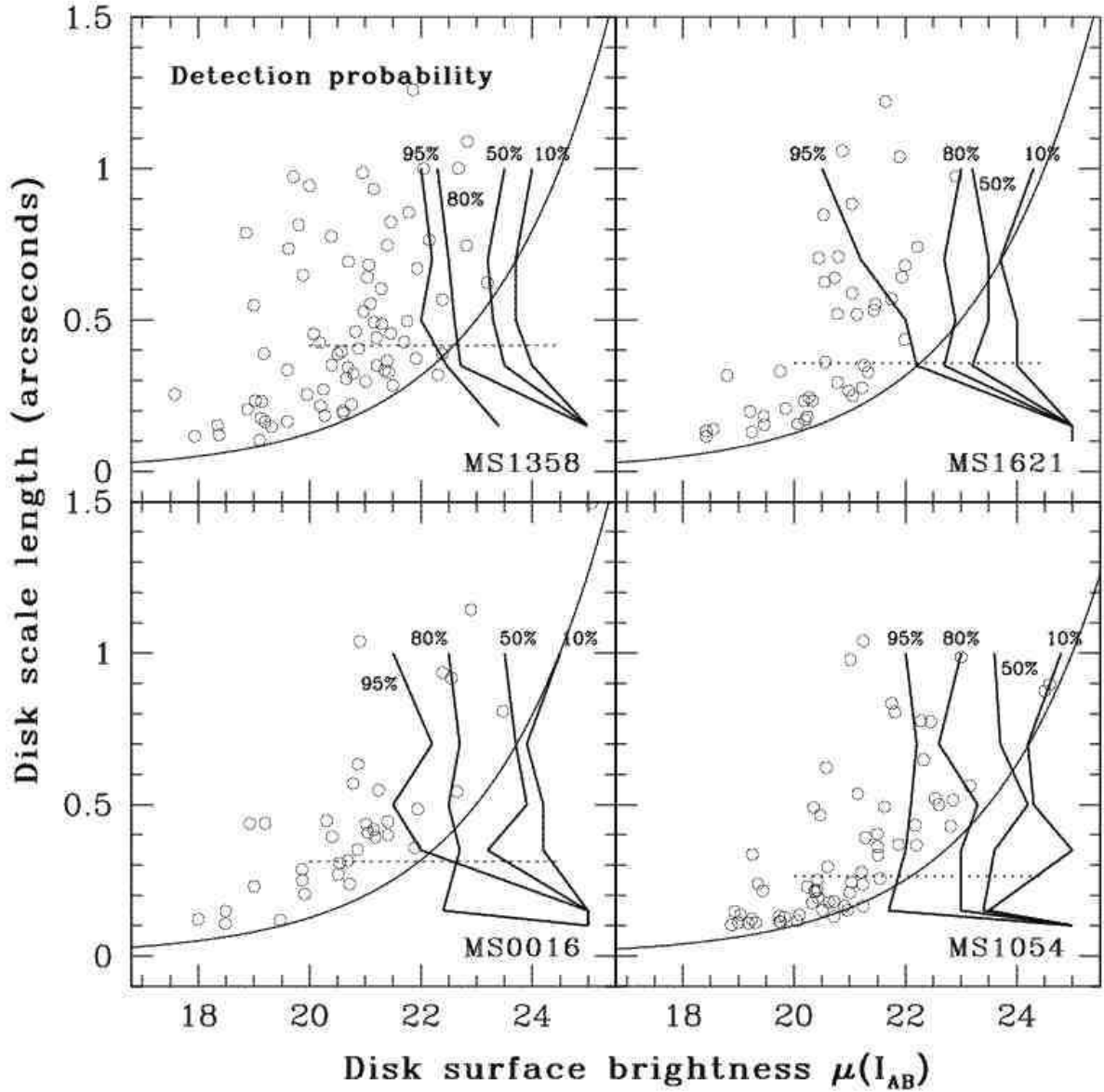}
\caption{Selection probabilities for pure disk galaxies in the four clusters.
\label{select}}
\end{figure}

 Figure \ref{select} shows the detection probability as percentages in the plane of
observed disk central surface brightness and disk scale length in arcseconds. This figure shows
the results for pure disk galaxies only. For a galaxy with a disk of a given
size and surface brightness the probability of detection is increased with the
addition of a bulge component.
 The lines on Figure \ref{select} show the nominal limiting magnitudes (F814W(AB)=23.0 
for MS1054 and F814W(AB)=22.5 for the other clusters) and the horizontal lines indicate
the angular size that corresponds to a scale length of 2 kiloparsecs. The simulation of the
detection process for disk galaxies produced the probability contours at 10\%, 50\%, 80\%, and
95\%. 

 Brighter than the nominal limiting magnitude the galaxy detection probability is very high 
(roughly 95\% or better) for all disks with sizes of 2 kiloparsec or smaller. At larger sizes
the samples are incomplete for low-surface brightness galaxies even though they may have
integrated brightness higher than the nominal limiting magnitude. For example, in these clusters,
disk galaxies with central surface brightness of $\mu(I_{AB})\sim23.5$ and sizes of 0.8 arcseconds
have a probability of being detected that is approximately 50\%. In all of the clusters the
region of large, low surface-brightness disks is sparsely populated. It is not possible to know
from these data and this detection procedure whether disk galaxies are present in those regions 
but remain undetected or are simply not present.
Therefore, we can make unbiased
comparisons of the properties of the disk galaxy populations among these
clusters only for scale lengths smaller than 2 kpc and at apparent brightness
higher than the nominal limiting magnitude.

\subsection{Spectroscopy and Astrometry}

In order to get redshifts and other spectroscopic information from different
surveys \citep[]{Fish98, CNOC3, CNOC4, CNOC6} , accurate coordinates were 
 required for each galaxy.  From the known pixel coordinates and the
WCS positions of each image, approximate coordinates are calculated using
the METRIC task of the STSDAS package in IRAF.  This task corrects for geometric
distortions specific to the WFPC2 on HST. When compared to stars  
that appear on the HST frames, systematic errors are frequently observed in the coordinates
from the METRIC task due to errors in the HST guide star positions. 
To correct these errors, each frame was searched for USNO catalog \citep[]{USNO} stars
and a mean systematic offset was derived. The systematic errors should be reduced to a few
tenths of an arcsecond (RMS) compared to the USNO catalog system.
All the positions given in this paper are these corrected coordinates.

\section{MORPHOLOGY \label{morphology}}

A total of $\sim4700$ galaxies were found by SExtractor above the initial detection 
threshold for each cluster (F814W$<24$ for MS1358 and MS1621,
and F814W$<24.5$ for MS0016 and MS1054).
 Quantitative morphological measurements are made for these galaxies by fitting parametric
models to the luminosity profiles \citep[]{Schade95,Schade96,marleau}.  The use
of parametric models is motivated by the fact that the galaxy profiles are of similar
size to the instrumental point-spread-function and need to be corrected for this effect 
and for sampling.  A set of commonly-accepted models is adopted here. Exponential and
de Vaucouleurs profiles are used. One advantage of this quantitative approach to morphology
is that a number of measurements (size, surface brightness, ratio of bulge luminosity to total
luminosity) are derived from the images rather than simply a morphological class. A second
advantage is that this scheme can be used to deduce evolutionary phenomena by comparing
nearby and distant galaxies, if one is careful to use data that are truly comparable. Clearly,
HST resolution is very valuable at $z > 0.5$ but moderate aperture telescopes with moderate
ground-based seeing can produce imaging of nearby galaxies that has similar signal-to-noise
ratio at a given luminosity and similar physical resolution, in kiloparsecs, to HST. These
local samples form ideal comparison datasets for HST. Further, if both the nearby and
distant datasets are analyzed in a uniform manner, it is reasonable to believe that accurate 
evolutionary information might result from such comparisons, despite the fact that it
is well-known that most galaxies show deviations, of varying degree, from the ideal models
adopted here.

 A disadvantage of the present approach is that comparisons with studies that are
based upon visual classification (e.g. \citet[]{dress97,vandenbergh}) are difficult.
For example, an intermediate galaxy might be defined (as it is here) as having a ratio
of bulge-to-total luminosity $0.4 < B/T < 0.8$. Such a classification undoubtedly includes
galaxies which would be defined as Sa, S0, and E galaxies using visual classification
methods. This is a strong motivation to draw conclusions from the comparison
of samples that have all been analyzed in a consistent way. Comparisons with work 
where visual classification is an important feature need to be done cautiously.

\subsection{Modeling}

For a single galaxy, the images produced with both filters were fitted
simultaneously. The size, ellipticity, and orientation
of each component of the galaxy is assumed to be identical in both filters.
The relative amplitudes of the bulge and disk components are allowed to
vary independently so that the model may reproduce a galaxy whose bulge
is a different color from its disk. We chose to use the integrated color
of the galaxy to accomplish the k-corrections because the individual
component colors were very noisy relative to the integrated value and
we present only the integrated values in the catalogs.
``Postage stamp'' images of the galaxies in the object catalog were produced
by first measuring accurate coordinates for the
center of each galaxy, and by measuring  
the sky value.  Then, a square image centered on the new galaxy coordinates is
produced and the sky is subtracted from it.  The size of the box
is chosen according to the fitting radius of the galaxy, which is
calculated with parameters produced by SExtractor.

The first step of the fitting technique is to produce a symmetrized image,
in order to eliminate strong asymmetric features or close companions.
This is achieved by rotating the stamp by 180\degr , subtracting that rotated
frame from the original one then clipping at $2\sigma$ (where $\sigma$ is the
error in a given pixel accounting for sky noise and noise form the galaxy itself)
of this difference
image to leave only significant positive deviations from symmetry. Then this
clipped image is subtracted from the original image to
produce the symmetrized frame \citep[]{Schade95}. 
A point spread function is then made, as described in \citet[]{Schade96}. 

Symmetrized galaxies
from both filter images are then fitted with three different models:
a $r^{1/4}$ bulge, an exponential disk and an hybrid model where the
total luminosity is divided between a bulge and a disk components.  For
each model, the best fit is determined.  The optimization process 
is done by varying the main morphological parameters: the size of the model, the axial ratio of the 
ellipse and its position angle, and by calculating the value of the $\chi^2$ \citep[]{Schade95}.  In addition, 
the best fit is then submitted to a ``trend test''
\citep[]{stat84} to look for systematic errors with radius 
left behind when the modeled galaxy is subtracted from the original image.
This second test is necessary since the $\chi^{2}$ does not account for the ordering
of the pixels: it computes the distance between each pixel and the sky level as if
they were all independent.  Thus, a mediocre fit can yield a fairly good
result at the $\chi^{2}$ test.  In this second test, pixels are ordered by radius
from the centroid of the galaxy.  The radial coordinate for this test is computed in the
frame where each model is circular (in other words the ellipticity is taken into account
when computing a "radial" variable for this test). Using this radially-ordered sequence,
each pixel is compared with its neighbors to look for any
 systematic of trends.  This is done by computing the mean square successive difference,
$\Delta^{2}$,
\begin{displaymath}
\Delta^{2}=\sum{\left( \begin{array}{c} x_{i}-x_{i+1} \end{array} \right) ^{2} / \left( \begin{array}{c} n-1 \end{array} \right)}
\end{displaymath}
where $(x_{i}-x_{i+1})$ is the flux difference between two successive pixels in the chain, and $n$ is the 
total number of pixels.
If there are no trends present, that difference is expected to be about
$2\sigma$.

Each galaxy has then six values associated with it that will come to play in
determining its type (bulge, disk, or bulge and disk) : the $\chi^{2}$ and the
trend statistic results for the best fit of each of the three models. Note that the
values of the statistics are computed over all of the pixels in the two bands that
were used in each fit.

\subsection{Classification}

\subsubsection{Visual Inspection}

Our strategy was to use visual inspection of the
fits to create a training set that would then be applied uniformly to the results
of the entire fitted dataset.
A large number of galaxies were examined visually together with the
fitted models and the residuals of those fits. The residual images were the most helpful at assigning a type to the galaxies. 
This procedure also provided a subjective estimate of the effect of
nearby galaxies on the fits (which should have been largely removed by the symmetrization process)
and also an estimate of the frequency of failed fits. Visual inspection 
made it possible to estimate the
magnitude, for each cluster, at which it became impossible to make a meaningful discrimination
between models.

\begin{figure}
\plotone{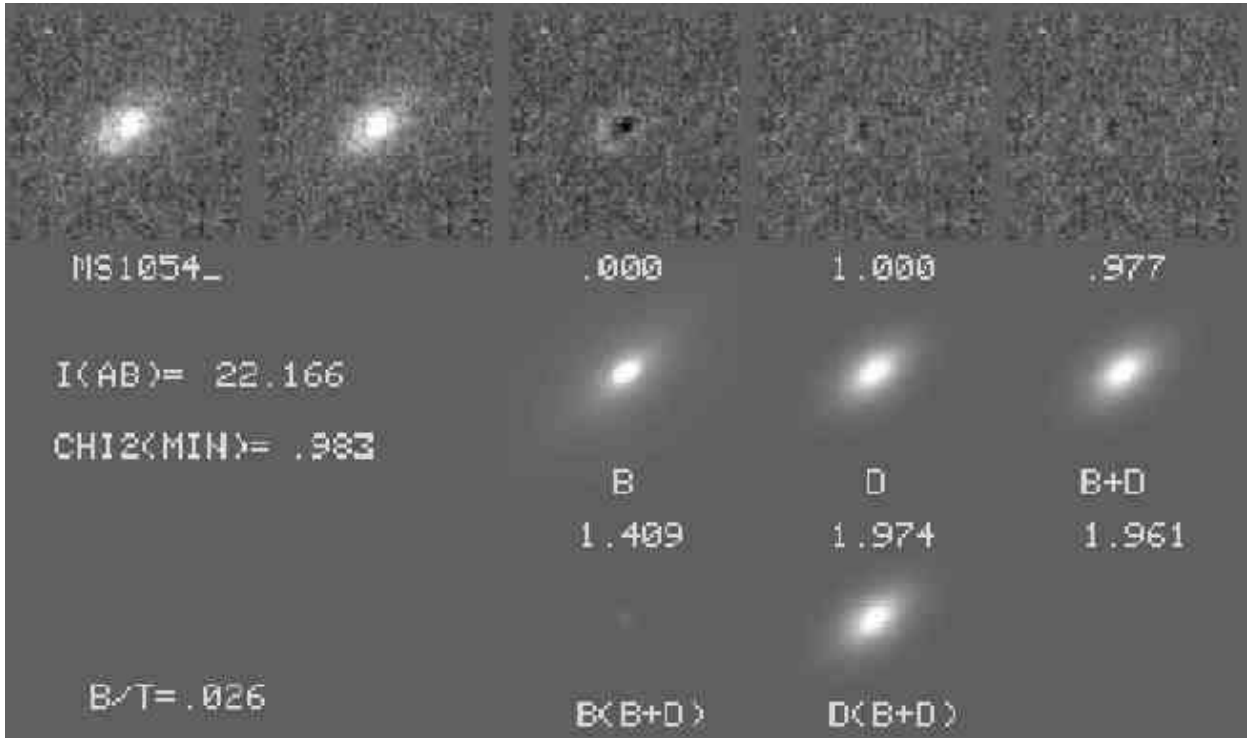}
\caption{The image display used to evaluate the results of the model-fitting procedure. The top
row (left-to-right) shows the original image, the "symmetrized" image, and the residuals after
the best-fitting bulge,disk, and bulge-plus-disk model is subtracted from the original image. More
details are in the text.
\label{screen}}
\end{figure}

 The process of deciding on class assignment (bulge, disk, or bulge-plus-disk) included access to a file
of the parameters and the statistics of the fit. These played an important role in the
decisions that were made. We show in Figure \ref{screen} the type of display that was used in
the classification.  The five images in the top row of Figure \ref{screen} are (left-to-right): 
the original
image, the symmetrized image, the residual after the best-fit bulge has been
subtracted from the original image, the residual after the best-fit disk has
been subtracted and the residual after the best-fit bulge-plus-disk model
has been subtracted. The numbers below the residual images are the F-test
probabilities that the fit is as good as the best fit (the best fit always
has a probability of 1).  The 3 images in the row second from the top are the 
best-fit bulge, disk,
and bulge-plus-disk model images. The value of the trend statistics
is below those images. The 2 images on the bottom row are the bulge component
of the bulge-plus-disk model and the disk component of the bulge-plus-disk model.
Other information is also marked on the display. In this example 
the disk model is the best fit although the
two component (bulge-plus-disk) model cannot be rejected at even the 5\% level.
It makes no difference in this case which of the two acceptable models is
adopted because the fitted disk parameters are nearly identical and the bulge
component is negligible, as confirmed by the images. The pure bulge model is
rejected at a high level of significance.

The most obvious way to choose the best fit is to take the fit
with the smallest value of $\chi^2$. The reason for rejecting this approach is that the 
models which represent the best fit from a statistical standpoint sometimes contain
components that are "un-physical" in the sense that they have components that are
effectively trying to fit faint residuals of nearby neighbors that have not been fully
removed by the symmetrizing process, artifacts of some kind, or even low-level errors
in the sky-level estimate. This happens most frequently with bulge-plus-disk models where
a legitimate (physical) component is accompanied by a second component which is
unrealistic and often has very low surface brightness. A visual examination quickly 
identifies such components but the statistics cannot directly do so. It is important
to note that our final catalog will certainly contain some errors of this type and 
any analysis based on the catalog needs to be sensitive to that fact.

When the quality of the fit for different models was too similar to 
effectively differentiate between a single
component model (either bulge or disk) and the bulge-plus-disk model, the galaxy was assigned the
type of the simpler model (disk or bulge).  In the cases where it was impossible to visually
discriminate between the bulge and the disk models, the type assigned was ``uncertain''.
Galaxies with such type helped determine the faint magnitude limit for the sample. 
Initially, all galaxies with observed magnitudes (in the AB system)
$F814W<24$ for MS1358 and MS1621,
and $F814W<24.5$ for MS0016 and MS1054 were fitted.  However, the types assigned to galaxies fainter
than a magnitude of 23 are mostly ``uncertain''.  This limit of F814W$<23$ corresponds
exactly to that used by \citet[]{dress97} when visually classifying galaxies from
similar images.

Approximately 1100 galaxies ($\sim 20\%$ of the data set) were inspected and classified as bulge, 
disk, bulge-plus-disk,
or uncertain and the fit quality was recorded. 
The galaxies classified are a representative sample of the whole catalog, with galaxies
out of each cluster and proportionally representing the whole magnitude range.

Most of the visual inspection was done by a single observer (AS), but a subset of 
150 galaxies was repeated 
by a second observer (DS), and the two sets of classes were compared.  Agreement occurred in 72\%  of
the cases, with every disparity resulting from galaxies that fell past the faint magnitude
threshold established above, or for which a single component model and the combined
one were practically equivalent.  There were no cases where one observer selected the bulge model
when the other one preferred the disk.  

\subsubsection{Automated Classification}

This evaluation of the fitted models for 20\% of the sample was meant to produce a training set
that could be used to perform an objective automated classification on the entire sample.  
The goal of
this process was to assign a fit type to every galaxy by comparing them to those
that had been visually classified.  The obvious interest of this method is to allow for a 
uniform classification of every galaxy, based on criteria set by the observer
when the training set is built.

Galaxies were removed if they were classified by eye 
as ``uncertain'' or had fits of poor quality, for example if a nearby bright 
object contaminated the image.  That left a training set of
415 galaxies, from all the clusters and spanning the whole range of magnitudes,
although with fewer galaxies at the faint end of the distribution after the removal of the ``uncertain''
class.

 The automated classification of the galaxies was made using the seven-dimensional
space defined by the probabilities derived from the $\chi^{2}$ test (3 values, one for each fit), 
from the trend statistics (3 values again), and by the F814W magnitude. Each galaxy was selected
and assigned the same classification (bulge, disk, or hybrid) as 
its nearest neighbor in this parameter space. This approach was implemented because
of its potential for rejecting un-physical fits. For example, it would be possible to
enforce a tendency to reject very low surface brightness bulge components if a
reasonable disk was already present. In fact, the nearest neighbor procedure is
very general. In the end we chose not to use any of the physical parameters of the
galaxies (size, surface brightness) except apparent magnitude in the automated
classification. 

 An optimization procedure was used to determine the normalization of the axes of the 
parameter space which produced the largest success rate for the training set where
success is defined as agreement of the automated classifier with the visually-selected
class. The success rate is not sensitive to the exact values of the normalization
factors and we set them to a ratio of 
of 2:4:1 for the $\chi^{2}$, trend statistics, and magnitude,
respectively.  With these values, the success rate for the computer based classification of
the galaxies from the training set is 85\% .  The discrepancies occur for the fainter
galaxies, or for the ones for which a single component model and the hybrid model were
almost identical.  In the former case, there will be a cut off in magnitude eliminating
these uncertain classifications.  In the latter, we notice that the classification
process would select the combined model when the observer would prefer the single
component one, but since the morphological parameters for both models are very similar in
such cases, the effect is
very minor.  

For each cluster, a sample of the galaxies that had not been visually classified were 
examined to verify the automated classification procedure.  No major disagreements were
noticed, but the rate of error with magnitude
rises significantly above F814W(AB)=23.  This
reinforces the faint limit of 23 established by observing at what point many galaxy types
were ``uncertain''.  There is also an increase of the error rate at the bright end of the distribution.
This effect is caused by the large resolved structures in these luminous galaxies, which are fitted
with less efficiency.  
However, these bright galaxies are not numerous
and are not likely to be at cluster redshifts.  Thus, the increase of errors at the 
two ends of the magnitude distribution will not significantly affect the results.

 The distributions of morphological parameters in our catalogs are robust in the sense that
they depend only weakly on the exact method that is used to facilitate the classification.
For example, we could have used a much more direct method of choosing the best fit: accept the
fit which produced the smallest value of $\chi^2$. If this had been done then we would have
produced the same classification as our nearest-neighbor procedure for 78.4\% of the training
set galaxies or the same classification as our visual classes for 78.9\% of the galaxies. As noted, we
chose not to do this because we could produce a slightly higher success rate (85\%) with the
nearest-neighbor method and we could reject some, not all, "un-physical" best-fit models. 
The relationship between the classification procedure we used and the statistics of the fit is made
more clear by examining how many real disagreements exist between the statistics of the
fit and our ultimate classification. A real disagreement is produced when our classification
is rejected by the F-test in favor of a classification whose fit is better from a statistical
point of view. Using this definition only 2.9\% of our training set galaxies represent disagreements 
with the statistics in the sense that our automated classification chooses a model whose fit is
worse than the best-fit model at the 95\% confidence level.
A comparison of the visual classification and the statistics yields an identical rate
of agreement. In other words, the choices made in our classification are supported by the
statistics of the fits in the overwhelming majority of cases.

\subsection{Morphological Parameters} 

The catalog contains a total of 1642 galaxies.  Cluster MS1358 has 672 galaxies, MS1621
has 258, MS0016 has 331 and MS1054 has 381.
Tables \ref{tab0016}, \ref{tab1054}, \ref{tab1358} and \ref{tab1621} present example of the results of the modeling and classification (the full tables are available only on-line).  The name of the galaxy in
column (1) contains the following information: name of the exposure, chip where the galaxy was found,
$x$ and $y$ pixel coordinates on that chip.  An asterisk put after the name indicates that
the galaxy has been statistically identified as a field galaxy.
Columns (2) and (3) give the coordinates of each galaxy. 
The observed magnitudes are given in column (4), and are the magnitudes in the reddest filter for each galaxy, F814W.
Color is given
in column (5) and corresponds to F814W-F555W for clusters MS0016 and MS1621, F814W-F606W for MS1054 and
exposures A through I, L and M of MS1358.  Exposures J and K of MS1358 are F814W-F450W.
All magnitudes are in the AB system.   Note that the magnitudes are calculated from the best-fit model
parameters and are thus effectively integrated to infinite radius.
In the next column (except for MS1054) are presented 
the redshift of the galaxies for which they were available \citep[]{Fish98, CNOC3, CNOC4, CNOC6}.

The last five columns of the table present the morphologic parameters.  In column (6) is $B/T$, which is
the ratio of light in the bulge component of the galaxy and the total luminosity.  If $B/T=0$, the galaxy
is a pure disk, and the last four columns present the properties of this disk, and if $B/T=1$ 
columns (7) through (10) list the quantitative
description of this bulge.  Finally, every galaxy with $0<B/T<1$ is represented by two lines in the table.
The first line gives its disk properties, and the second line lists the bulge parameters.
 In column (7), $\mu$ is the central surface brightness,
the scale length is H in (8), AR is the axial ratio (the ratio of the length of the minor
axis to the major axis of the galaxy) and is presented in column (9), and the last column gives the position angle
of the galaxy.  The position angle is the angle between the major axis of the ellipse and the $x$ pixel coordinate
axis on the original images.  The errors presented in columns (8) and (9) are mostly 
reliable for the single component models, and are given in the hybrid model case 
to indicate the average precision of the measurements.  

\subsubsection{Photometric bias and the symmetrizing process}

 After the fit has converged its parameters represent our best estimate of the
shape of the galaxy luminosity profile. Given that shape, the normalization of the
model is determined by

\begin{displaymath}
A={{\sum_i{ {O_i P_i} \over {\sigma_i^2 }  }} \over {\sum_j{ {P_j^2}
\over {\sigma_j^2 }
  }}}
\end{displaymath}

 where $O_i$ is the observed counts  in the $i^{th}$ pixel and $P_i$ and
$\sigma_i$ are the values of the model and the noise at that pixel.
 The normalization $A$ is the total number of counts in the model if 
$\sum_j{P_j}=1$. This relation can be derived by minimizing $\chi^2$ under
the assumption that the model is represented by the product of a shape
(the $P_j$'s) and a normalizing factor.

 As described earlier, the original image is rotated, subtracted and 
the difference image is clipped
at $2\sigma$ to create an asymmetric image composed of only the significant 
positive departures from symmetry. This image is subtracted from the original image.
Statistically, there is some flux removed from the image even if it is symmetric
because of the noise in the image. The size of the effect depends on the signal-to-noise
ratio and the most severe effect occurs at the
background limit (that is, when background noise from sky and read-noise is the dominant noise
source). Given a gaussian background noise distribution with a 
specific root-mean-square (r.m.s.) deviation or
$\sigma$ the clipping will leave behind 2.25\% of the pixels and these will carry,
on average,
a signal of 0.054 times $\sigma$ (background) for each pixel in the aperture. This signal
is subtracted from the image and thus produces a bias in the measured flux. This bias
is in the sense of producing less flux in the symmetric image compared to the
original image.

 The size of the effect depends strongly on the size of the photometric aperture 
 and we estimate the size of the effect using values of signal and noise from our actual data
for MS1358 where the galaxies are larger than in the more distant clusters. The larger size
translates into larger photometric apertures and thus more background noise and
a larger value of the potential bias. 
For the
F814W (I-band) images in MS1358 we compute a bias of about 3.5\%  at $I(AB)=22$ using 
an aperture with a diameter of 3 arcseconds (the bias would be 6\% at the limit of $I(AB)=22.5$ or
1.4\% at $I(AB)=21.5$). However, as noted above,
aperture photometry is not used to normalize our models.
The approach we use optimizes the signal-to-noise ratio by weighting the pixels according to how
much galaxy signal they contain. This reduces the effective number of pixels that contribute
to the noise therefore it also reduces the size of the bias. For a disk galaxy 
at $I(AB)=22$ with a scale length
of 0.5 arcseconds the bias is reduced by a factor of 0.57 to about 2\% and for 
scale lengths of 0.25 and 0.10 arcseconds the bias is reduced to 0.8\% and 0.2\% respectively. 
For bulge galaxies of a given half-light radius the effect is small because the galaxies
are more compact than disks. For example, at $I(AB)=22$ a bulge with half-light radius of 0.5 arcseconds
will have a bias estimated to be 0.5\%. So for
typical galaxies the effect is small. Still, the existence
of the bias and its variation with galaxy size is an effect to note with caution.
For faint large galaxies ($\sim 1$ arcsecond) the effect could still be 3-4\%. But the effect is small
in MS1358 and smaller still for our more distant clusters.

 Simulations of several thousand galaxies (typically 250 galaxies per run) 
were done to see if we could detect any bias in the
fitting of symmetrized images compared to fitting the original images. The exercise highlighted
the reason that the symmetrizing process was important: fits 
(particularly two-component fits)  to the original images are plagued
with problems due to neighboring stars and galaxies. For this reason, the 
results that follow were computed after excluding
5-10\% of the poorest fits. For disk galaxies with simulated scale lengths ($h$) of 0.5 arcseconds and magnitudes
of $I(AB)=22.5$ we recovered values of $h=0.52\pm0.15$ and $I(AB)=22.55\pm0.4$ from the original images
and $h=0.50\pm0.14$ and $I(AB)=22.62\pm0.4$ from the symmetrized images. The quoted errors are the dispersions
in the recovered values.
At $I(AB)=22.5$ and $h=0.5$ arcseconds these galaxies have poor signal-to-noise ratio which explains
their large errors.
For moderate size disks with $h=0.25$ arcseconds we recovered
$h=0.27\pm0.09$ and $I(AB)=22.48\pm0.24$ and $h=0.24\pm0.03$ and $I(AB)=22.57\pm0.2$ from the original
and symmetrized images. For very small disks ($h=0.10$ arcsceonds) at $I(AB)=22.5$ we recovered
values of $h=0.10\pm0.006$ and $I(AB)=22.54\pm0.06$ and $h=0.10\pm0.005$ and $I(AB)=22.51\pm0.2$ from the original
and symmetrized images. Any bias due to the symmetrizing process
is very small compared to the other sources of error, for example
crowding by neighbors, sky subtraction, PSF uncertainty, and centering errors.

 If the signal-to-noise ratios of the two images used to produce colors differ
significantly from one another, it would be possible to produce a color bias from the symmetrizing procedure.
This possibility was examined with the particular data from this paper and was found to be at
roughly the 0.5\% level when 3 arcsecond diameter apertures were used. The color bias
would be smaller in our fitting procedure as noted above and is thus a negligible
source of error.

 The referee for this paper made the interesting suggestion that it might be possible to
produce an artificial reddening of galaxies if blue star-forming regions which were 
distributed asymmetrically were removed by the symmetrizing process. Clearly the effect would
be expected to be largest at low redshift where the regions are best resolved. The most direct
way to address this issue is to perform aperture photometry (thus bypassing any dependence
on the model-fitting process) on the original images and compare that photometry to the same
measurements on the symmetrized images. We did this with 95 of the larger
disk-dominated galaxies in the field of MS1358
without regard for whether these had been classified as cluster members or field galaxies.
It was found that the offset in the mean color between the original images and the symmetrized images in
MS1358 was $0.014\pm0.016$ magnitudes in F814W-F606W. If a single outlier was removed then the offset
became statistically significant ($0.027\pm0.011$ magnitudes). If the offset were being produced
by the removal of star-forming regions by the symmetrizing process then blue galaxies would be
expected to show a stronger effect than red galaxies. 
A comparison of the blue and red halves of the sample 
yielded offsets of $0.033\pm 0.013$ and $0.02\pm0.02$ magnitudes respectively. So the
effect is small and is not significantly larger for the blue galaxies which should be forming stars
more vigorously. A small set of large disk galaxies is available in MS1358 with F450W and F814W
observations and this set should be more sensitive to the effect of star-forming regions
because of the bluer filter measurement. For this
set of 15 galaxies we find a reddening of $0.003\pm0.07$ magnitudes ($0.056\pm0.04$ if the worst
outlier is removed). Again, this fails to provide support for the suggestion that the symmetrizing
process is biasing the colors of galaxies by removing star-forming regions. This conclusion
is supported by a visual inspection of images and symmetrized images along with the fit residuals for
50 of the larger galaxies in both MS1358 and MS1054. There
are very few cases where there the effect of the symmetrizing process might plausibly produce a 
significant change in the structure and color of the galaxies.

\subsubsection{Errors on the Morphological Parameters}

 The errors on single-component models (bulge or disk) are given as output from the fitting
software and are generally reliable (e.g., \citet[]{crampton02}). Errors on multi-component fits suffer from
the correlation of errors in the bulge and disk components which are assumed to be concentric. Because
of these error correlations the errors from the fitting software may not be reliable when fitting multi-component
models. \citet[]{Schade99} used fits to multiple observations of the same galaxies to estimate the errors
in the morphological analysis.
The evaluation of the errors can be confirmed through simulations (e.g. \citet[]{Schade96}). 
For the present paper, several hundred simulations of galaxies spanning the range of morphological
parameters and signal-to-noise ratios were done. These confirm that the scatter from all sources of
random error for galaxies of the size and surface brightness that this paper deals with are typically in
the range of 10\% in the scale length or half-light radius 
and 10-20\% in surface brightness. The errors can be worse for large,
low-surface brightness galaxies where the detection is also problematic.

\section{RESULTS \label{results}}

\subsection{Classification Reliability}

\subsubsection{Definition of the Morphology Classes}

The galaxy types are
defined as follows: disk galaxies are taken to be the ones with $B/T\leqslant0.4$, and bulges those for which
$B/T\geqslant0.8$.  The galaxies with ratios in the range of 0.4 to 0.8 are said to be intermediate.  This class
is likely to include all of the S0 galaxies, as well as other bulge dominated structures that still possess
a disk (early type Sa galaxies, for example).  The comparison between these results and other work should
therefore be done cautiously.  But as will be shown in the next paragraphs, the classification still allows one to
recreate classical results as accurately as a visual classification with Hubble type classes would do.  

 A small correction based on the individual colors of the disk and bulge components
is applied to the $B/T$ values to reduce them to rest-frame $B$-band.

\subsubsection{The Color-Magnitude Relation \label{sectioncmag}}

\begin{figure}
\plotone{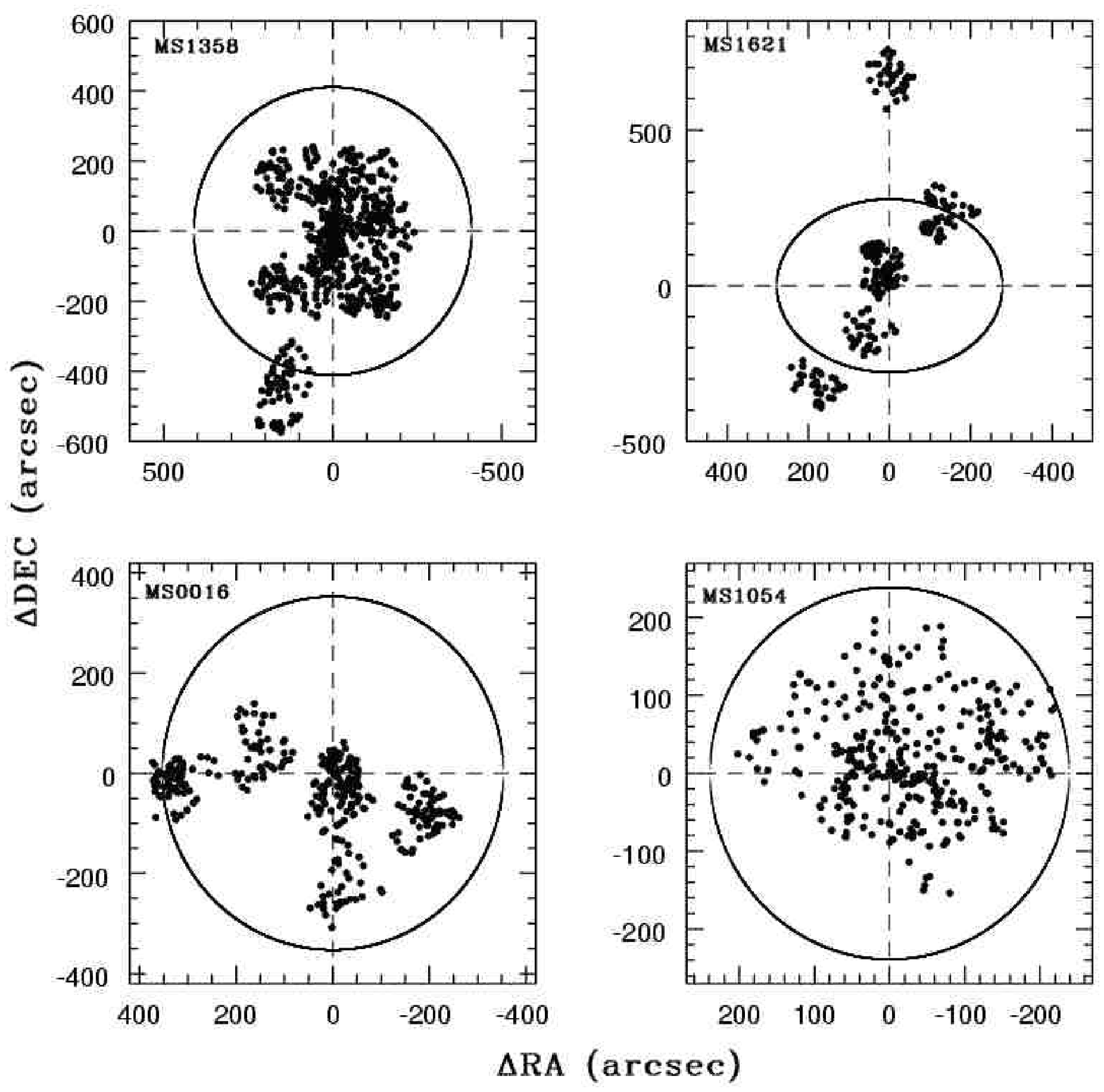}
\caption{Spatial distribution of the galaxies in the sample.
The lines plotted on each frame 
represent $r_{200}$, the characteristic radius of each cluster.  \label{radec}}
\end{figure}

\begin{figure}
\epsscale{0.9}
\plotone{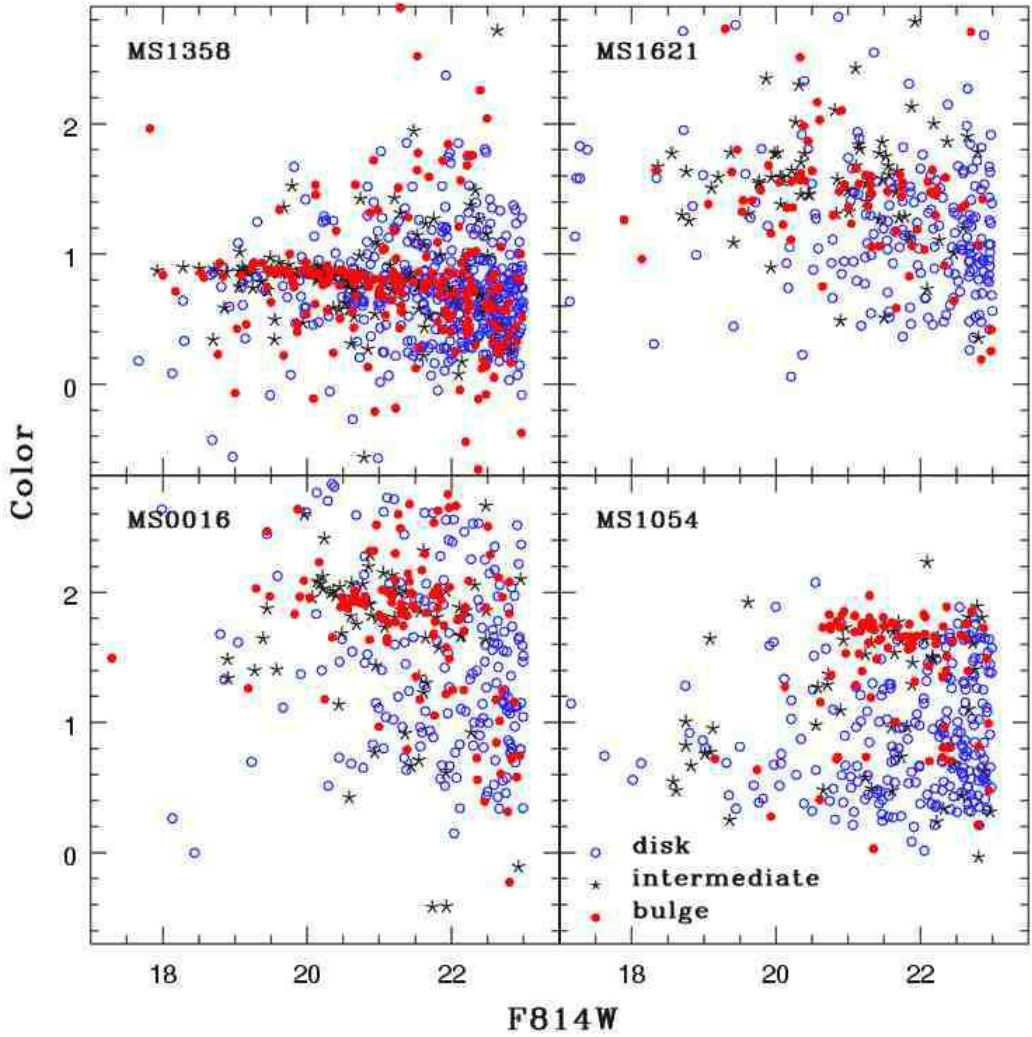}
\caption{Color-magnitude relation for each cluster.  Color is (F814W-F555W) for clusters MS1621 and MS0016
and (F814W-F606W) for MS1358 and MS1054, and the magnitude is the total observed in the F814W band.
\label{cmag}}
\end{figure}

 One test of the reliability of these galaxy classifications is an examination of the
color-magnitude relation. Bulge galaxies should be predominantly red and
exhibit a tight relation. Disk-dominated galaxies would be expected to show a
wider dispersion in color.
Figure \ref{cmag} presents the 
color-magnitude relations in the observed planes for the four clusters (of course
some field galaxies will also be present in these fields).
As expected, the distribution of colors differs between galaxies defined as bulges
and those defined as later type. The bulges tend to form a tighter sequence at 
redder colors than the later types, which cover a wide range in color.
This is especially clear for clusters MS1358 and MS1054 where 
the observed galaxies are concentrated in the center of
the clusters (cf. Figure \ref{radec}).

\begin{figure}
\plotone{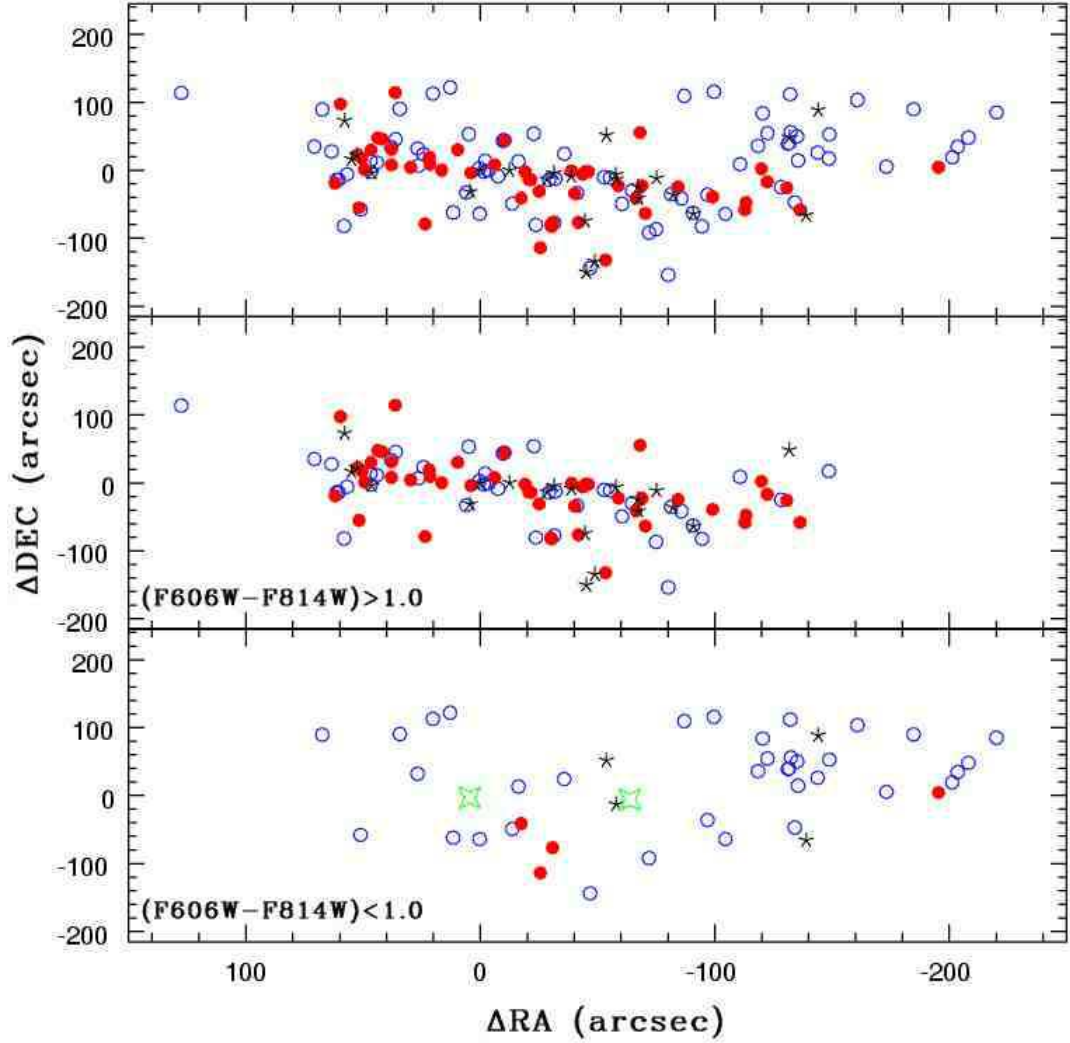}
\caption{Spatial distribution of the galaxies in MS1054 by galaxy type.  Open circles are disk galaxies,
filled circles bulge galaxies, and stars intermediate galaxies.  The top panel presents all
galaxies from the data set with F814W$\leqslant 23.0$, the bottom panels show the two groups identified on the color-magnitude diagram.
The open stars on the third panel are the centers of the high density regions observed in X-ray \citep[]{jeltema}. 
\label{c1054}}
\end{figure}

MS1054 shows a high concentration of faint blue disks, and an apparent gap in
color between the red sequence and this group of galaxies.  This suggests that there might be
some galaxy groups in the field that are unrelated to MS1054 or sub-clustering in MS1054 itself 
to produce that 
bimodal distribution. 
It has been shown using X-ray maps produced by \textit{ROSAT} and \textit{CHANDRA} that there is 
sub-clustering in MS1054 \citep[]{donahue,jeltema}.
To investigate the possibility that the bimodal distribution in the color-magnitude plane is 
caused by different groups being observed, the spatial distribution of the galaxies in MS1054 was examined.  
Looking at Figure \ref{cmag}, it appears that there are two distinct structures, with colors 
(F606W-F814W) larger and smaller than $\sim 1$. 
Figure \ref{c1054} shows the spatial distribution of these two subsets on
the bottom two panels, respectively.  As expected, the reddest galaxies (those with (F606W-F814W)$\leqslant1.0$) 
are more concentrated towards the center of the cluster as a result of the morphology-density relation.  
We then expect the bluer galaxies to be distributed around that central concentration.  This is what is 
shown in the bottom panel of Figure \ref{c1054}.  The distribution appears uniform 
and not to correlate with the two clumps observed in the X-ray. 
The central position of these concentrations, as given in 
\citet[]{jeltema}, are shown on the bottom panel of Figure \ref{c1054} as the open stars.   
The apparent presence
of a concentration of blue disks north and west of the center in that plot is directly related to the spatial distribution of the
observations made, as seen in Figure \ref{radec}.
Without redshifts to confirm the membership of these 
galaxies to the cluster, we will assume that we are only sampling a different population of the cluster 
by observing at outer radii and that these galaxies legitimately belong to the cluster. Note that a 
statistical correction will be made for field galaxy contamination.

\subsubsection{The Morphology-Radius Relation}

Another way of verifying the results of the classification, and the validity 
of the morphology classes, is to 
look at the morphology-clustercentric radius relation.
It has been shown that a radial dependence of galaxy morphology inside clusters exists \citep[]{oemler74, mel77}.  
It has also
been argued \citep[]{dress80} that instead of projected radius, local density should be used as the independent
 variable, since
it takes account of any sub-clustering or irregularities in the distribution of galaxies in the cluster, which was 
supported by the observation of a universal relation in local clusters between morphology and density.  Recently, a debate 
has been made as to which of the two relations is more fundamental (e.g. \citet[]{whit93}, \citet[]{dress97}, \citet[]{dominguez}).  Since we are looking at 
the morphology-radius only to confirm the validity of our classification, we will not present the 
morphology-density as well, nor 
try to answer that question.

\begin{figure}
\plotone{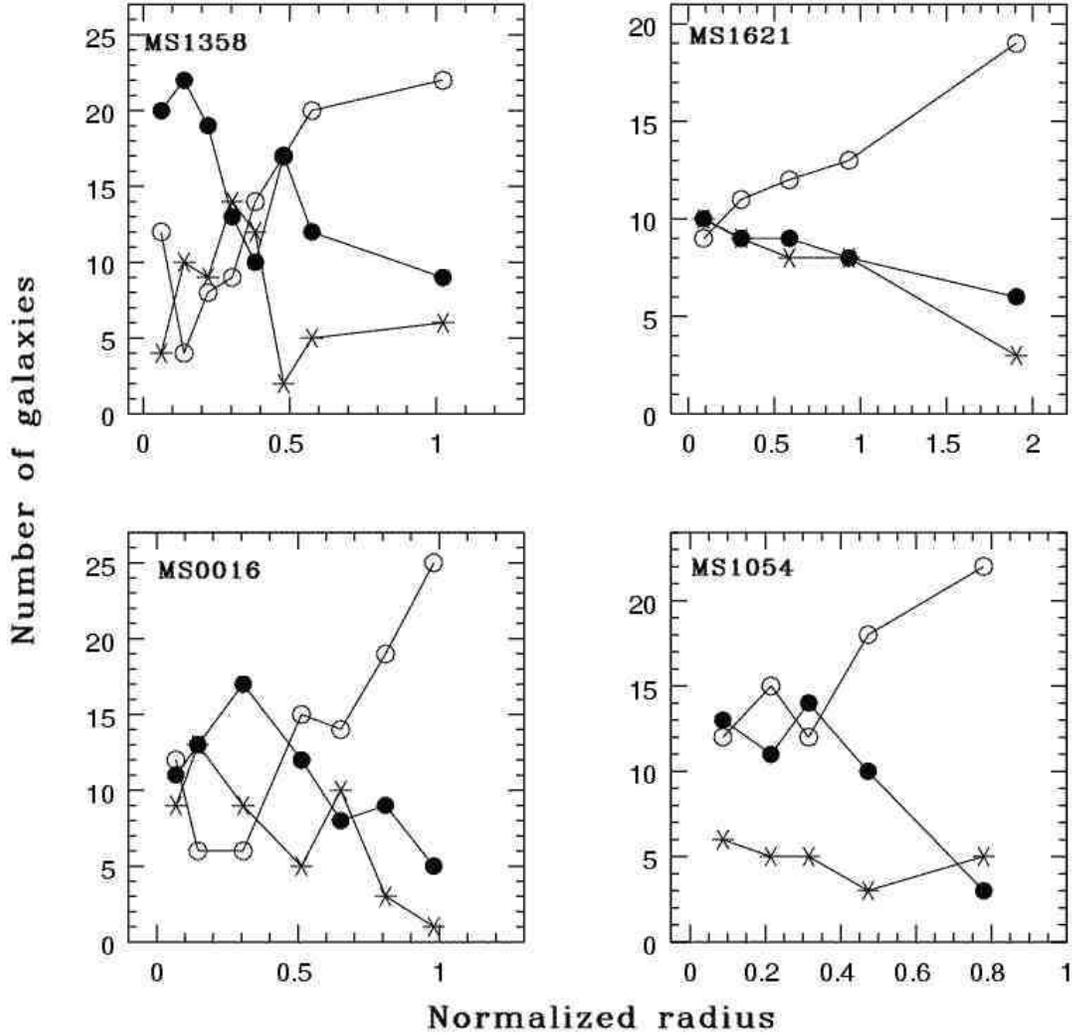}
\caption{Morphology-radius relation for each cluster.  For each cluster, the radius bins all contain equal numbers of galaxies.  The number of galaxies of each type observed in that bin is normalized by that total. 
Disk galaxies 
($0 \leqslant  B/T \leqslant 0.4$) are represented by open circles, bulges ($0.8 \leqslant  B/T \leqslant  1$) by filled circles and intermediate galaxies ($0.4 <  B/T < 0.8$) by stars. 
\label{morphrad}}
\end{figure}

The morphology-radius relation for our four clusters is presented in Figure \ref{morphrad}.  In all cases, but to various extent, the expected trends are observed: an increase in the disk population with radius, and a decrease in the number of intermediate and bulge galaxies. 

\subsection{Corrections}

Rest-frame values are calculated to allow for comparison between clusters.  The observed F814W(AB) magnitudes are 
k-corrected using the galaxy color to choose among the spectral-energy distributions (SEDs) 
of \citet[]{coleman}. Note that AB magnitudes are used throughout. The procedure used, 
including the convolution of filter bands with the
SEDs is described in \citet[]{cfrs}.
The same tables and procedures used in that work were used here.
The chosen SED is then used to determine the correction from the observed wavelength
to rest-frame $B(AB)$-band and used together with the systemic redshift of the cluster to determine
rest-frame $B(AB)$ absolute magnitude.  The various colors (F814W-F606W,F814W-F555W,F814W-F450W) 
are transformed into rest-frame $(U-V)(AB)$ (thereafter, $(U-V)_{0}(AB)$) using the same SED-choosing procedure.
For the galaxies that were assigned a hybrid model, 
k-corrections are obtained separately for the bulge and 
the disk from the color of each individual component.  Finally, disk scale-lengths and bulge half-light radii
are computed in units of kpc and surface brightnesses with restframe $B(AB)$ magnitudes.
All rest-frame values were calculated using $H_\circ=70$ km sec$^{-1}$ Mpc$^{-1}$,
$\Omega_m=0.3$ and $\Omega_{\Lambda}=0.7$.

\subsubsection{Field Galaxy Contamination}

Redshifts are available for only $\sim15\%$ of the galaxies, so in order to have the largest sample possible, 
another means of defining cluster membership must be used. A statistical correction for field galaxy
contamination was applied to identify probable cluster members. It should be borne in mind, however, 
that such a correction works only in the case where the level of actual field galaxy contamination is 
typical, in a statistical sense. If
there were groups or clusters of galaxies in the same field as the X-ray identified clusters but
at a different redshift, then this process would fail to weed them out.

To perform this correction, images of other randomly-selected fields obtained with WFPC2 were retrieved. 
They include 23 frames from the Groth strip and 12 frames from the {\it Canada-France Redshift Survey} \citep[]{cfrs}. 
The images were reduced, processed, and their galaxies were
fitted and classified in exactly the same manner as the cluster images.  For a given pair of filters, 
the same number of frames 
from the field were processed as for the cluster sample; the use of a single frame to correct 
all the data could introduce systematic errors.  In order to avoid such errors, the subset of field galaxies 
used to do the correction was randomly selected and changed for each cluster frame.  

It is necessary in the field correction procedure to account for the different density of sources
on the frames. The field galaxies are assumed to be uniformly distributed whereas the cluster
galaxies are concentrated toward the cluster center. On the scale of an individual HST 
pointing the distribution of
galaxies appears to be clumpy rather than obeying a clear radial density gradient and this clumpiness
varies from cluster to cluster. It was decided that the scale represented by a single HST pointing 
(there are 5 or more in each cluster) was
sufficient to characterize the local density for the purpose of doing he field correction and
this had the advantage of avoiding the assumption of a very regular galaxy distribution which
is only true in an average sense. Each pointing was corrected individually.

The randomly selected galaxies for a given field were ``matched'' with the cluster galaxies.  The cluster galaxy that 
most closely resembles each field galaxy is  
marked as a ``statistically identified field galaxy'' and removed from the cluster sample.  The parameters 
used to do that matching are the observed magnitude, the color, and the $B/T$ ratio. The best match
was defined using the same type of procedure as applied for the classification. A space was constructed
and the nearest neighbor of each field galaxy in the cluster was removed.
Clusters MS1358, MS1621 and MS0016 were corrected up to a magnitude of $F814W(AB)=22.5$, and MS1054 up to $F814W(AB)=23.0$ . 

\begin{figure}
\plotone{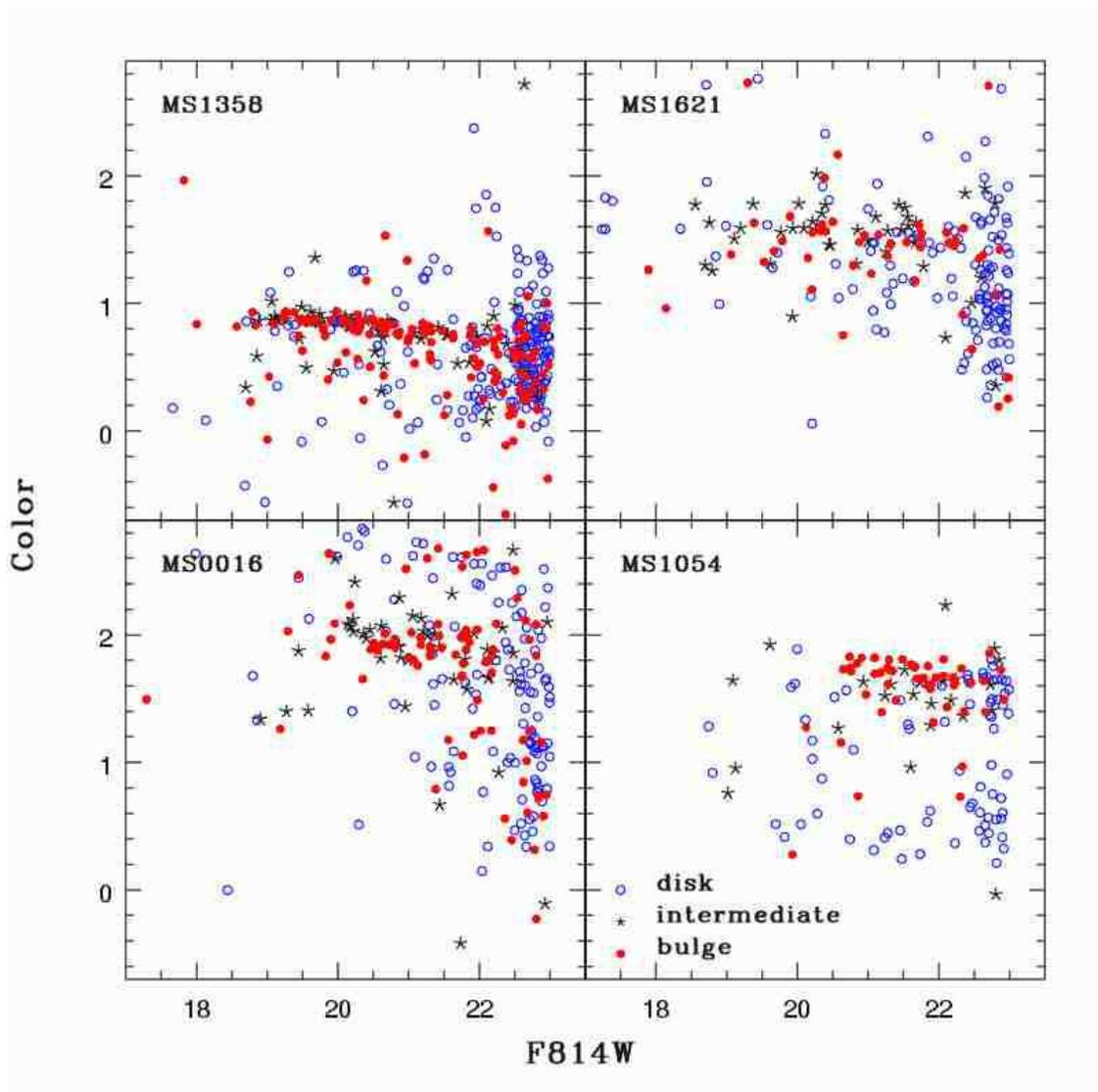}
\caption{Color-Magnitude relation, after the field contamination 
correction is applied.  Color is (F814W-F555W) for clusters MS1621 and MS0016
and (F814W-F606W) for MS1358 and MS1054.  Open circles represent disk galaxies, stars intermediate galaxies, and filled circles bulge-dominated galaxies.
 \label{fieldcorr}}
\end{figure}

The color-magnitude relation is used in Figure \ref{fieldcorr} to illustrate the effect of the 
field galaxy contamination correction. 
The red bulge sequence is relatively un-modified and many of the galaxies off the sequence
are identified statistically as interlopers from the field.
For MS1358, there was a large number of galaxies redder than the reddest 
ellipticals in the core of the cluster, which were believed to be either higher redshift background galaxies or 
very dusty foreground galaxies.  The correction identifies most of them as field galaxies.   
In MS1054, as discussed in \S \ref{sectioncmag} , there are two groups visible in the color-magnitude diagram,  
one which lies in the bluer region of the plot (c.f. Fig. \ref{cmag}). That ``blue group'' of galaxies is 
largely removed by the statistical correction process. 

Another means of verifying the results of the field galaxy correction 
is to compare the sample of field galaxies used 
to do the correction to the ones rejected from the cluster frames.  This comparison is done using one of the 
parameters that remained free during the matching process: the disk scale length.  The distribution of the 
sizes from the two samples were compared using a Kolmogoroff-Smirnoff test with the result that the
hypothesis that the two samples of scale lengths are drawn from the same distribution cannot be rejected
at the 5\% level (the K-S probability was 7\%). This indicates that the galaxies in the cluster field that
are designated as field galaxies are similar to the field galaxies used to do the corrections. We have forced
them to be similar in terms of color, magnitude, and bulge-to-total luminosity and, after doing so, we find
that the scale length distributions are not distinguishable.

\subsubsection{Normalization by Mass}

 It is impossible to do a reasonable comparison of the number counts of galaxies between the four
clusters directly because
the clusters vary in mass and because the completeness and the radial coverage of the sampling
varies enormously from cluster to cluster (as shown in Figure \ref{radec}). 
In order to correct for this problem, a normalization factor was computed based on the fraction
of cluster mass that was sampled by the fields observed in the present study.
The calculations take into account the mass of each cluster, the field of view, and the density profile of the 
clusters, and are made under the assumption that the cluster mass is a good predictor of the total number 
of galaxies it contains.  In other words, we are assuming that the efficiency of galaxy formation is identical
in these four clusters. Further, we implicitly assume that the efficiency of production of each galaxy type is
constant across clusters. 

In a study based on the CNOC clusters, \citet[]{carlberg97} have shown that the galaxy number density profile 
is proportional to the cluster mass profile, at least over a given range in 
radius ($0.1\leqslant r/r_{200}\leqslant1.5$).  
Therefore, by selecting an appropriate density profile and 
integrating it over the field of view, a good estimate of the mass that has been sampled can be made. This
can be converted into an expected number of galaxies under the assumptions stated above.
In the study by \citet[]{carlberg97}, it is shown that the function 
$\rho(r)\propto r^{-1}(r+a_{\nu})^{-2}$ is adequate to describe 
the mass distribution of 14 of the 16 CNOC clusters, including the three studied here (this model is similar to the one presented by \citet[]{navarro}).  Therefore, it 
will be used to represent the three-dimensional spatial density profile of the clusters, with a value for the scale 
radius, $a_{\nu}$, of 0.27 as suggested by the best fits of \citet[]{carlberg97}.  

To compute the normalization factor, the assumption that the clusters are spheres containing their galaxies within 
$r=1.5r_{200}$ is made.  First, $\rho(r)$ was integrated to project it in the plane.  This produces $\phi(r)$, the 
surface density which is then normalized by the integral of $\rho(r)$ over the whole sphere.  
This normalized surface density is then integrated over the entire field of view of the observations.  The fraction 
of the cluster observed is then known, and multiplying it by the cluster mass 
(as presented in Table \ref{tab1}) gives the normalization factor that we seek.  

The results are sensitive to the cluster masses. 
Estimates of the mass of MS1054 have been made using various techniques: X-ray luminosity \citep[]{donahue,jeltema}, 
weak lensing \citep[]{hoekstra}, and observed velocity dispersion \citep[]{tran}.  
The value calculated with this last 
technique is the one adopted here, since it is very similar to that used to evaluate 
the masses of the CNOC clusters \citep[]{mass}. 
The values of the normalization factor are presented in Table \ref{tabnorm}.  

\subsection{Disk Galaxy Surface Brightness Distributions \label{analysis}}

In this section, the distribution of disk surface brightnesses is investigated in the four clusters.
 It is important to note that these galaxies are identified as ``disk''
galaxies solely on the basis of the light profiles, in the sense that exponential profiles provide better 
representations of their luminosity distributions than are provided by de Vaucouleurs profiles. 

\subsubsection{Magnitude Selection}

\begin{figure}
\epsscale{0.65}
\plotone{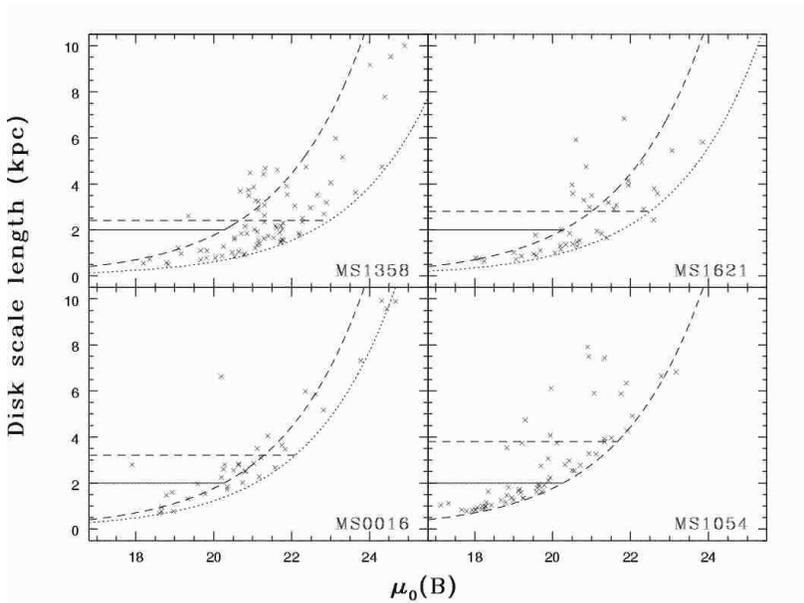}
\caption{Relation between the disk scale length and rest-frame B disk surface brightness for the 4 clusters.
In each panel the selection line due to the limiting magnitude ($F814W(AB)<22.5$ for MS1358, MS1621 and 
MS0016) is indicated by a dotted, curved line, and the corresponding line for MS1054 ($F814W(AB)<23.0$) is
shown on all panels as a dashed curve. A horizontal dashed line indicates the angular size of 0.5 arcseconds
where incompleteness starts to become significant. We also indicate 2 kpc by a solid line. Below 2 kpc
the samples of disk galaxies are complete to the nominal limiting magnitude. Below 2 kpc and to the left
of the limiting magnitude dashed line for MS1054 all of the clusters are complete so that the
disk populations can be directly compared.
\label{hdsb}}
\end{figure}

Figure \ref{hdsb} shows the relationship between the scale length and the rest-frame
B-band central surface brightness of the disks
in those galaxies with $B/T\leq0.4$ and with magnitudes brighter than 
F814W(AB)=22.5 (23.0 for MS1054).
At each cluster redshift, the cutoff in observed magnitude translates into a cutoff in luminosity 
(modified slightly by variation of the K-correction with color). 
Disk central surface brightness ($\mu_{0 disk}$)
depends on both disk scale length ($h_{kpc}$) and absolute B magnitude ($M_{B disk}$) as :
\begin{displaymath}
\mu_{0 disk}=M_{B disk}+2.5\log{\left(  2\pi \right)}+5\log{\left( h_{\rm kpc} \right)}+36.565  
\end{displaymath}
The dotted lines represent these limiting luminosities in terms of size and surface brightness.
 Figure \ref{hdsb} also indicates on each frame the magnitude cutoff for MS1054. 

 The analysis of the sensitivity of our source detection procedure to disk galaxies shows that
the probability of detecting pure disk galaxies in all of the clusters is high ($> 95\%$) brighter
than the nominal limiting magnitudes only for those galaxies with scale lengths smaller than
2.0 kiloparsecs in all clusters.
Therefore the
region of the diagrams shown in Figure \ref{hdsb} to the left of the magnitude selection lines
and with scale lengths smaller than 2.0 kpc is fairly sampled in all of the clusters. That is,
fair comparisons of 
the size-surface brightness distributions can be made only in this region. In practice we
limit our consideration to disk galaxies with $h < 2$ kpc.
 The most striking feature of Figure \ref{hdsb} is the increase with redshift in the number of small ($h < 2$ kpc),
high surface brightness disk galaxies. This effect could be produced in a number of ways.
This could be an actual evolutionary effect. Or it could be that the populations of disk
galaxies in all of the clusters are similar but that the sampling rate of the population in
MS1054 is much higher than in the other clusters so that the observed number is larger although
the underlying distributions are identical. If the clusters do not represent a family with a
common evolutionary history then the difference in the number of small disk galaxies may
not be an evolutionary effect but merely an indication of the range of characteristics 
of the galaxy populations in rich clusters.

 In the next section, an attempt will be made to address the issue of the different sampling rates
in the four clusters by normalizing the counts on the basis of the sampled mass in each
observation.
This will be based on the conjecture that disk galaxy formation is equally efficient in all of these clusters.
Another way of saying this is that a given quantity of total mass (dark plus baryonic) will produce a
given quantity of disk galaxy mass. The uncertainty of this conjecture is obvious and the fact that
the disk galaxy populations may be in different phases of their evolution in different clusters
is a further caveat.

 Nevertheless, we will proceed to investigate the difference in the properties of these 
galaxies under the assumptions that we have stated above. The simplest pair of models to describe
the change in the size-surface brightness distribution is a) a shift in size of the entire distribution
with no change in surface brightness, and b) a shift in surface brightness of the population with no
change in size. 

\subsubsection{Pure Disk Size Evolution}

 Pure size evolution of disk galaxies (by definition with no change in surface brightness)
is a way to model changes
in the statistical distributions that are observed in these samples. But this model is difficult
to link in a simple way to a physical scenario for the evolution of individual galaxies because some
change in star formation rate is likely to occur over time and passive evolution of the stellar population
will change the disk surface brightness even if the star formation rate is maintained a some constant
level. It is conceivable that a changing star-formation rate could balance an aging population that is
fading but it seems unlikely because it would require an interesting degree of fine-tuning if it were to
be observed over a population of galaxies.

In general, evolving disks are expected to grow in size as they add more mass 
(e.g., \citet[]{MoWhite}).
 In the present case
we observe many small galaxies at a given central surface brightness in MS1054 at $z=0.83$ which are absent in the lower
redshift clusters. The way for them to escape detection in these lower redshift clusters is for them to
evolve to smaller sizes (at a given surface brightness) as time moves forward. This is directly contrary to
a simple model of disk growth. Therefore, we have reason to be skeptical of the pure size evolution
as a physical model for evolution for individual galaxies. Nevertheless, we will test pure size evolution
as a model for the evolution of the distributions of size and central surface brightness.

To do this test, the size distribution of each cluster is shifted (in size only) until 
the ``equivalent expected number'' of galaxies above the selection line is the same as the
observations of MS1054. This means
that the ratio of observed numbers of galaxies above the selection line is equal to the ratio of
the normalizing factors which correct for cluster mass and the spatial coverage of the observations,
as calculated in \S 4.2.2  (the ``equivalent expected number'' for each cluster is given in column 2 
of Table \ref{tablum}).  In other words, the whole galaxy sample for a given cluster is shifted in 
size until the expected number of galaxies appears in the region of the size-surface brightness plane 
where the small bright disks are found.  This region is the one represented in Figure \ref{hdsb}.
The shifts in the disk scale length distributions required are given in the third 
column of Table \ref{tablum}.

\begin{figure}
\plotone{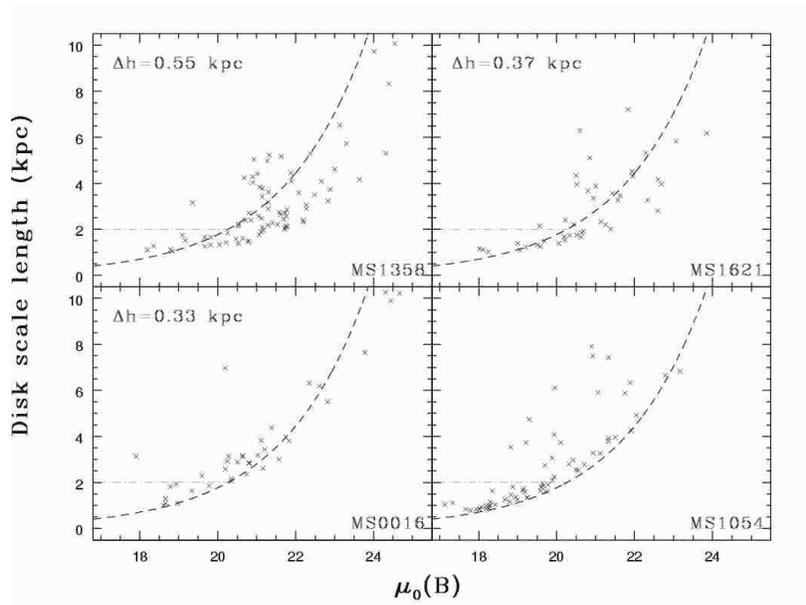}
\caption{Testing the size evolution hypothesis: the size distribution of each cluster is shifted so that the 
same expected number of galaxies is present above the selection line and below the 2 kpc horizontal line.  The model fails in the case of MS1358 and MS0016 where it cannot provide the number of expected disks.  Surface brightness is in the rest-frame B(AB). \label{delh}} 
\end{figure}

A problem with this test appears immediately: a shift in size is unable to provide the expected number of 
small bright disks for MS1358 and MS0016. We can only shift the distributions in size until the
selection line due to limiting magnitude in each cluster coincides with the corresponding selection
line in MS1054. Beyond this shift we have no sensitivity to smaller galaxies in MS1358, MS1621, and
MS0016.
The shift given in Table \ref{tablum} is the shift that gives the largest number of small 
bright disks and is roughly the largest shift we can do for MS1358 and MS0016.
This problem is not due to the poorer sampling of these clusters since the normalization 
technique used to calculate the expected numbers removes the effects of the different areas sampled. 
The shifted distributions are plotted in Figure \ref{delh}.  In that plot, the lines are the same as in 
Figure \ref{hdsb}, i.e., the magnitude cutoff of MS1054 and the 2 kpc scale length cut off.  

\subsubsection{Pure Disk Surface Brightness Evolution}

 The limiting size at which detection of disk galaxies is complete varies from cluster to cluster as
given above. At this limiting size there is a limiting central surface brightness imposed by the magnitude limit.
The region of the size-surface brightness plane that
is complete for all of the clusters has a limiting size of 2.0 kpc and and at this scale length each
cluster has its own limiting surface brightness. For MS1358, MS1621, MS0016, and MS1054 this
limiting surface brightness is roughly 22.6, 21.8, 21.1, and 20.3 respectively. Thus we can shift each cluster
population by, at most, the difference between its limiting surface brightness at 2 kpc and that of
MS1054. If we shift it more than this amount then we are no longer comparing complete regions of the
size-surface brightness plane.

\begin{figure}
\plotone{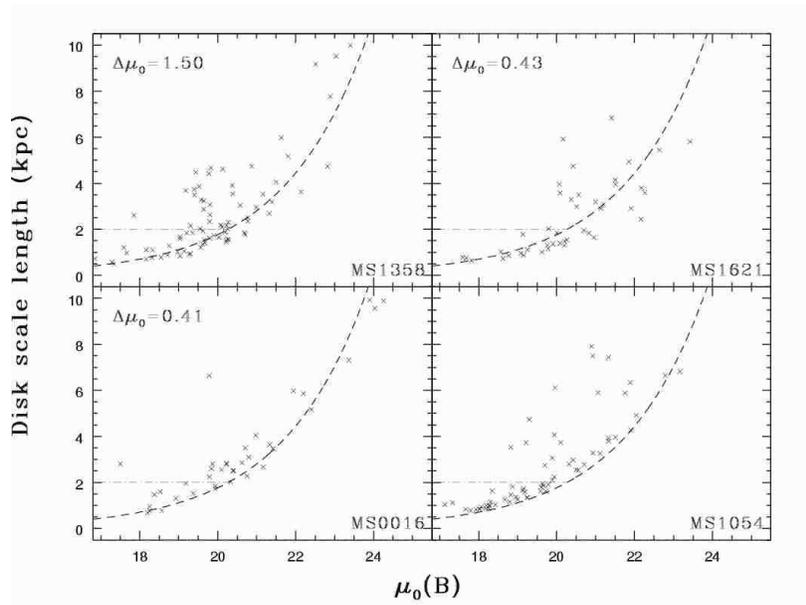}
\caption{Estimation of the shift in the surface brightness distribution.  
The surface brightness distribution has been shifted until the same expected 
number of galaxies was above the selection line and below the 2 kpc horizontal line.  
The shift in surface brightness required is indicated in each frame.  
Surface brightness is in the rest-frame B(AB). \label{delsb}} 
\end{figure}

 The simplest model that is suggested by these data is a uniform change in surface brightness experienced
by the entire disk galaxy population. To test this model the population is shifted in surface brightness only,
subject to the constraint above, namely, that we are careful to compare only well-sampled regions of
the size-surface brightness plane. The shift produces an acceptable model when the observed number in
the well-sampled region of the clusters agree with the expected number of galaxies predicted from
the MS1054 data using the mass normalization to account for cluster mass and variation of our sampling.
The results are shown in Figure \ref{delsb}. The 95\% confidence intervals for these shifts
were estimated using procedures described by \citet[]{gehrels}.

 The number of disks in the region where they can confidently be compared (scale length smaller that 2 kpc 
and surface brightness high enough to be detected in MS1054)
is different in MS1358 and MS1054 at greater than the 99.9\% confidence level. The galaxy populations differ
by more than the difference in sampled mass would indicate. The shifts that are required in order to bring the
number of observed galaxies in our comparison region into line with the number predicted by the mass
normalization factor (together with the number in MS1054) are given in Table \ref{tablum}. At the 95\% confidence
level only MS1358 differs, to a degree that is statistically significant, from MS1054 although the best-fit 
shifts in surface brightness are always non-zero.

 The comparison of MS1358 and MS1054 indicates that a shift of $1.50^{+0.26}_{-0.80}$ magnitudes 
(95\% confidence level) in surface brightness produces agreement in the number of small disks and,
furthermore, the median scale lengths and dispersion  
in the resultant samples agree reasonably well (the values are shown in Table \ref{tablum}). The statistics for the other clusters are poor because
of small numbers of galaxies.

 A simple model such as pure evolution of the disk population in surface brightness
is unlikely to provide a complete explanation of the observations. Nevertheless, with
the statistics in hand on these four clusters, such a model explains the main features
of the size-surface brightness distributions of small disks in these four clusters. In particular,
it models the discrepancy in the small disk population between MS1358 at $z=0.3$ and
MS1054 at $z=0.83$.

\begin{figure}
\epsscale{0.75}
\plotone{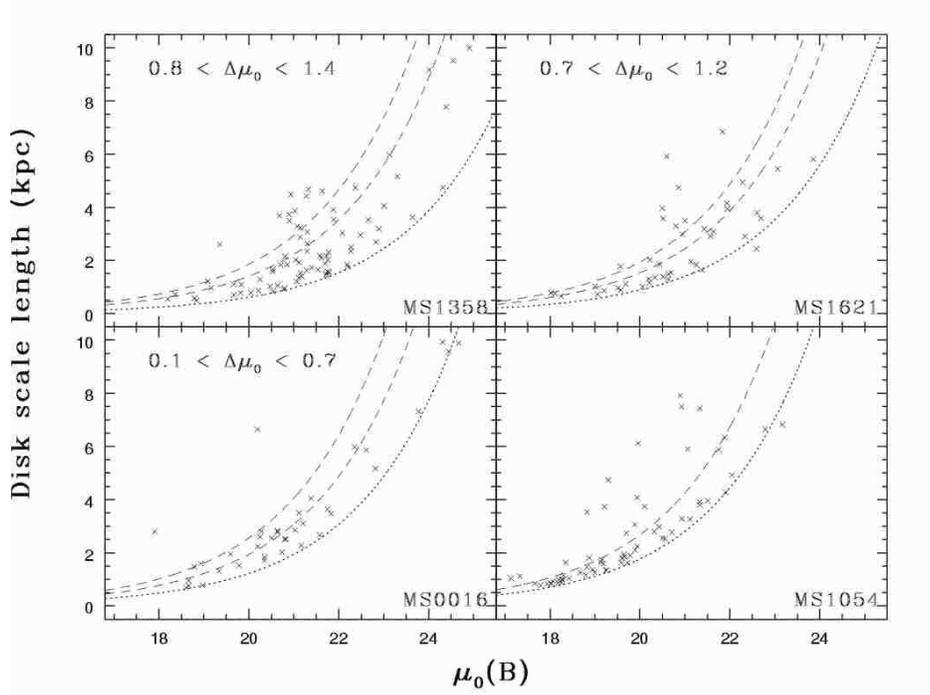}
\caption{Estimation of the surface brightness enhancement with redshift.  Lower and higher estimates of the position of the upper limit of the distribution are plotted as the dashed lines.  The dotted line represents the faint magnitude limit for each cluster.  The ranges for the shifts in surface brightness from MS1054 are indicated in each frame. \label{env_minmax}}
\end{figure}

 Another way of estimating the apparent change in surface brightness of these galaxy populations
is by visual examination of the upper-left envelope to the distribution in the plane of
size and surface brightness. This envelope, where the maximum scale length at a given
surface brightness increases with fainter surface brightness may be explained
by the higher angular momentum of low surface brightness galaxies, which results in larger 
disk sizes for a given mass \citep[]{dalcanton}. Since the existence of
this envelope is, plausibly, an intrinsic physical property of the galaxies, 
tracking it through the different clusters 
allows one to get an estimate of the luminosity evolution of the galaxies.  On Figure \ref{env_minmax} are 
traced both the limiting magnitude of each cluster 
and the estimated position of the upper envelope, represented by constant magnitude lines on each frame.  
Both upper and lower limits for this upper envelope are plotted. 
This time, the $\Delta \mu_{0}$ represent the shift between the upper envelope of each cluster and the one of 
MS1054.  The shifts are simply estimated by eye.
The values are given in the last column of Table \ref{tablum}.
The ranges just overlap with those previously calculated in the cases of MS0016 and MS1621, 
but are somewhat smaller for MS1358.
There is reasonable agreement of this 
result with the previous estimation of the surface-brightness evolution that used a more rigorous
statistical procedure.

In summary, a simple model of the change in the distribution of small disk properties in the
plane of size and surface brightness is pure surface-brightness evolution. If interpreted physically
this implies a brightening of $\sim 1.5 $ magnitudes in the B band over the range $z=0.3$ to $z=0.8$.

\subsubsection{Can we discriminate between disks and bulges at small sizes?}

\begin{figure}
\epsscale{0.75}
\plotone{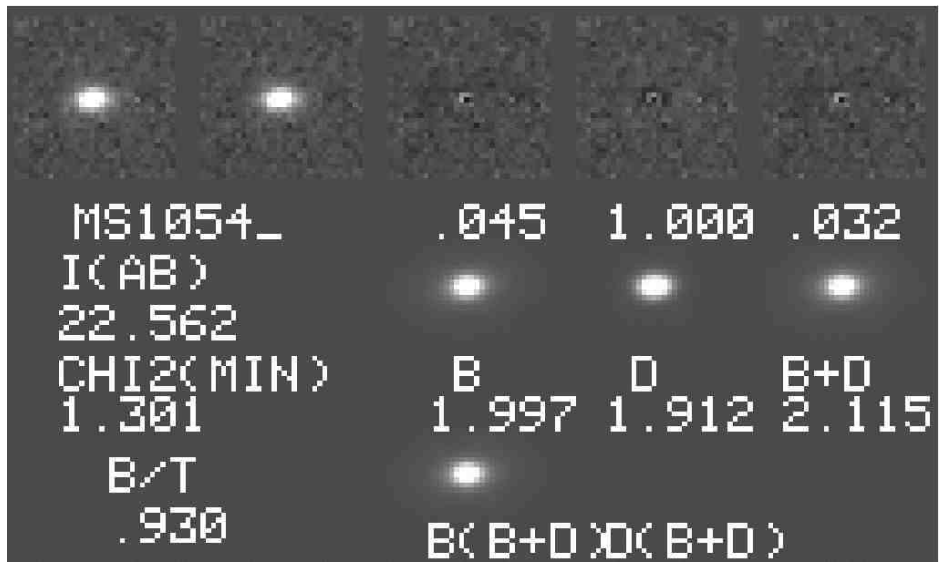}
\caption{An example of the fits to a very small galaxy with $I(AB)=22.56$ in MS1054. The arrangement of
the images in the top row are as discussed in Figure \ref{screen} and show that we can distinguish
statistically between bulge and disk models even where the derived disk scale length is 
very small ($h=0.108$ arcseconds). Note that the light profile, convolved with the point-spread
function can be traced over many pixels although the scale length is only slightly larger than
1 pixel.
\label{diskfit}}
\end{figure}

 The results of the preceding sections raises the question of whether we could be making the
error of classifying galaxies as disks when they are actually bulges. The frequency of such
errors might vary with redshift producing a spurious population of small disk galaxies at
high redshift. In Figure \ref{diskfit} we show that we can distinguish reliably between
disk and bulge models even for very small and faint galaxies in the field of MS1054.

 In order to quantify our success rates at distinguishing bulges and disks we simulated galaxies
including all sources of error and subjected them to the end-to-end fitting and classification
process. Note that the fits to actual data were done by simultaneously fitting the images in
two filters whereas these simulations were done using only a single image. The magnitude scales
were accurate so that the signal-to-noise was reproduced exactly for the I-band images in 
the simulations. This means that our simulations actually had lower signal-to-noise ratios than
the two-filter fits that were used for the real data. Therefore, these simulations produce results 
that are lower quality than our actual fits. This simplified the simulation procedures and is
sufficient to illustrate the effectiveness of the fitting process.

 The resolution in physical units is lowest in MS1054. Simulations were done of 800 disks with a scale length
of 0.10 arcseconds and these galaxies were added to actual MS1054 images in order to
include the effects of neighbors and crowding.
 The galaxies went through
the same detection and symmetrizing processes as the real data. Uncertainties in the point-spread function
and sky level were effectively included. The fits were processed through the automated classification
procedure. 

 Pure disk galaxy models with true (simulated) scale lengths of 0.1 arcseconds and 
$I(AB)=22.5$ were classified as pure disks in 90.2\% of the cases. In 91.8\%
of the cases the disk parameters were recovered effectively with the
galaxy classified as either pure disk or bulge-plus-disk. In the latter case
this implies a negligible bulge component although the class chosen was
bulge-plus-disk. In 3.4\% of the cases the simulated disks were classified
as pure bulges. In the remaining 4.8\% of the cases the fits failed
to converge to a reasonable result. 

The mis-classification as disks of galaxies that are actually bulges is a serious
concern. We simulated and fit approximately 800 pure bulge galaxies 
with half-light radii $R_e=0.167$ arcseconds and $I(AB)=22.5$. This corresponds to the
same half-light radius as a pure disk with scale length of 0.1 arcseconds.
We claim success (recovery as a pure bulge in 73\% of the cases or 
recovery as a bulge-dominated bulge-plus-disk galaxy
where the bulge parameters were well-determined in 23.9\% of the
cases) in a 96.9\% of the cases. Our end-to-end simulations
mistakenly classified a pure bulge as a disk in 1.3\% of the cases. The
remaining 1.8\% were fits which failed to converge.

 These simulations also yield estimates of the errors on our fitted parameters
by comparing the distributions of recovered parameter values to
the known input values. The errors do not decrease rapidly with increasing signal-
to-noise ratio indicating that they are dominated by small systematic errors 
such as errors in the point-spread function, sky level estimate, galaxy centering,
and other details of the processing procedure. For the smallest galaxies 
(half-light radius $\sim 0.2$ arcseconds) at $I(AB)=22.5$
the dispersion in both recovered size and magnitude hover around 10\% and grow
to about 15\% as the size doubles at a fixed luminosity. As the half-light
radius approaches 1 arcsecond the dispersion in recovered size approaches
30\% and the error in brightness is about 0.35 magnitudes.  There are not
many galaxies this large in the catalog. Systematic
errors are seen as failures of the mean recovered values to converge on the true value
even for large sample sizes but these systematics are always less than the
dispersions and the worst cases are for small galaxies where the systematics
can be 50-60\% as large as the random errors. For the majority of galaxies
in our catalog the total errors in size and brightness parameters are in the 
range of 10\% for smaller galaxies and 20\% or less for most of the
catalog galaxies. 

 In summary, these results indicate that we can reliably distinguish between bulges and disks
near the magnitude limit of our observations and at very small galaxy
sizes where the interesting results of our study fall.

\section{DISCUSSION \label{discussion}}

 The cluster MS1054 apparently contains a population of small, high-surface brightness disk-like galaxies that
are not seen in the other three clusters at lower redshift. 
These galaxies are 
referred to as ``disks'' because they are better fit by an exponential profile than by a de Vaucouleurs
$r^{1/4}$ law. 
These galaxies  are reminiscent of the high-surface brightness field ``disk'' populations detected by 
\citet[]{Schade95,Schade96}, and \citet[]{guzman96}.
Under the assumption that the observed differences in the
small disk-galaxy populations in these four clusters are due to evolution, then pure surface
brightness evolution describes the observations 
somewhat better than pure size evolution. Neither model is perfect and our ability to distinguish between them
is limited by the small numbers of galaxies involved in the comparison.
More complex models have not been investigated.

Using two independent techniques, the luminosity enhancement of the disk galaxies was estimated.  The 
method based on the detection of the upper limit of the distribution of galaxies in the size-surface 
brightness plane is free of any assumptions on the mass and population of the cluster.  It provides a range of values 
for the shift in surface brightness observed in the galaxy population from cluster to cluster.  The other technique, 
based on the cluster mass normalization factors, gives another estimate for each cluster. 

The agreement between the results of the two techniques is encouraging in terms of the validity of the counts normalization method.  To further test it, larger samples would be required such that statistics could be made with larger number.  As it was seen with these clusters, the larger the data set, the easier it is to get accurate results when trying to find the ``equivalent expected number'' of galaxies, as done in section \ref{analysis}.  One must also keep in mind the assumptions that were made, mainly that the number of disks observed in a cluster is directly proportional to the total mass of the cluster, and that the galaxy number density profile is proportional to the mass profile.  Even though this second assumption is supported by previous work \citep[]{carlberg97}, the first one albeit reasonable has yet to be proven.

Because of all the assumptions made, the result on the luminosity evolution of the galaxies needs to be considered carefully, in the context  
of previous work.  \citet[]{Schade96b} have shown that disk galaxies,  
both in the field and in clusters, have a mean surface brightness increased by $\sim1$ magnitude at $z=0.55$ 
compared to local clusters.  Similar conclusions have also been reached in other works (e.g. \citet[]{forbes96,roche}). 
\citet{mallen99} studied the population of small bright disks and showed that this rapidly-evolving population has sizes and velocity widths typical of irregular galaxies in the local Universe but are typically $\sim 2$ magnitudes more luminous than
local irregulars. 

The most obvious caveat accompanying our conclusions about the evolution of small disk galaxies
is that the cluster membership of these galaxies is uncertain. Redshifts are not available for all galaxies. Corrections for
field galaxy contamination have been made but these corrections apply only to typical fields and it may be that
there are atypical levels of contamination by non-cluster members. In other words, we may be seeing an effect
that is not only unrelated to the cluster environment but is erroneous because the cluster redshifts have been
assumed for all cluster ``members'' and these redshifts would be incorrect if membership were incorrectly assigned.
Redshifts would make these conclusions far more secure.

\begin{figure}
\plotone{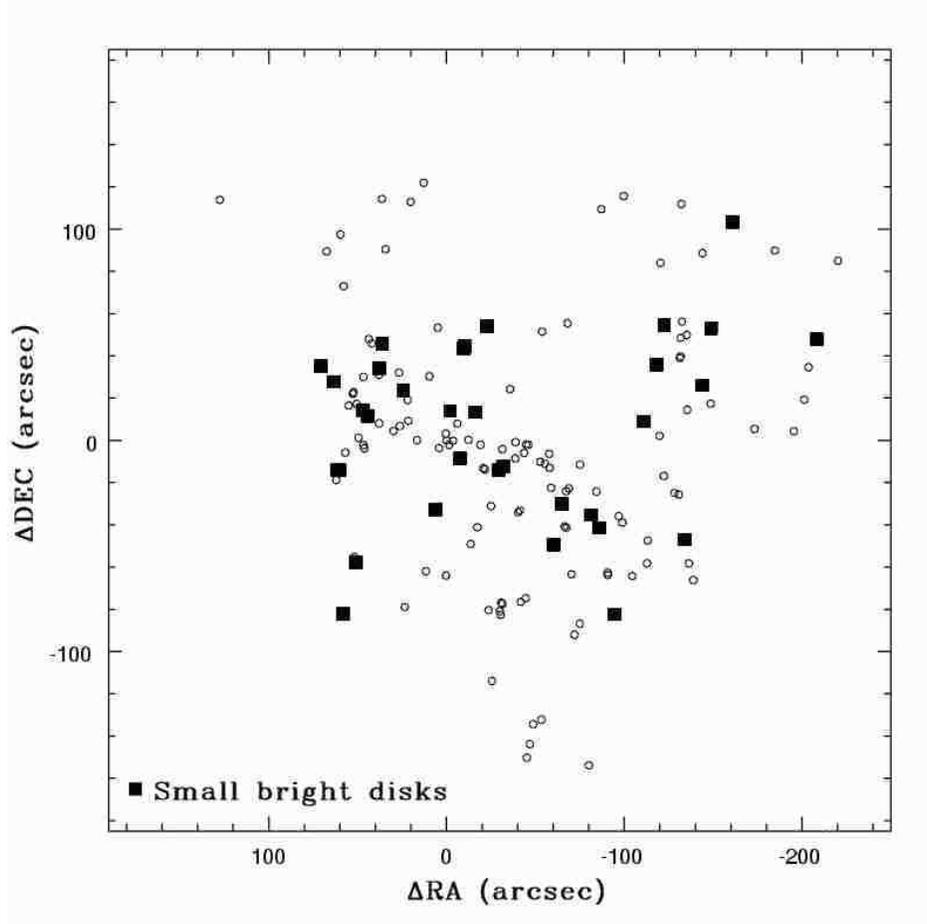}
\caption{Distribution of the small ($h \le 2$ kpc), high surface brightness disks in MS1054.  The filled squares represent these small disks and the open circles the remaining galaxies of the sample. 
\label{radec1054} }
\end{figure}

To further rule out the possibility that the small bright disks observed in MS1054 are contaminants, their spatial distribution is observed.  The small bright disks are not concentrated in any region of the cluster, as evidenced by Figure \ref{radec1054}, such as it 
would be expected if they were members of a group projected in the field of MS1054, or of a subgroup falling into the cluster.  The small bright disks of all clusters were visually examined. They all appear to form one population of compact and regular disks.  The fact that these galaxies are observed in all clusters, even though there are many more in MS1054, is another argument to assume that they are not observed in MS1054 because of a field-of-view contamination.

\section{SUMMARY AND CONCLUSION}
A catalog of $\sim1600$ galaxies from the rich clusters MS0016, MS1054, MS1358, and MS1621 is presented.  
It shows positions, photometry, and quantitative morphology obtained through modeling of the galaxies.  Each galaxy was fitted with three different models (bulge, disk, bulge and disk), and assigned a type based on 
the comparison of the best fit of each model (as evaluated by the $\chi^{2}$ optimization).  The galaxies
were then classified by both an observer for a subset, an by a computer script for all the data set.  
The classification produces the morphology measurements: the bulge-to-total luminosity ratio, the 
scale-lengths, axial ratios, position angles and surface brightnesses of both the bulge and disk components 
of the galaxies.

The validity of these techniques was tested by comparing the results they produce to well known 
properties of cluster galaxies.
First of all, the color-magnitude relation made with observed values is examined.  The galaxies assigned a ``bulge'' 
type are concentrated in the region of the red sequence, a tight distribution in the color-magnitude 
plane.  This is in agreement 
with the known fact that the red sequence is formed by the ellipticals in the center of the cluster.  
The disk galaxies generally have bluer colors, and the scatter increases with fainter magnitudes.
Secondly, the morphology gradients observed in the clusters through the morphology-radius  
relation with the morphology classes defined in this paper are those expected: the number of 
disk galaxies increases with radius, while the intermediate and bulge galaxies are more numerous 
in the inner regions of the cluster.  

The catalog of measurements was applied to the study of luminosity evolution of disk galaxies.  A population of small ($h<2$kpc), high surface brightness disks is observed in MS1054.  
Some of these disks are also present in the other clusters, but in much smaller numbers.  Two models were tested to account for the presence of the small bright disks in MS1054: the first model considers a size evolution with redshift, and the second a luminosity evolution.  The size evolution model is not successful at describing the data.  The luminosity evolution model does a good job, albeit not perfect, at explaining the differences in population with redshift.  As a result, a luminosity enhancement of $\sim 1.5$ magnitude from $z=0.3$ to $z=0.8$ was observed.

\clearpage

\begin{deluxetable}{llllccccl}
\tabletypesize{\scriptsize}
\tablecaption{Cluster properties \label{tab1}}
\tablehead{
\colhead{Cluster} & \colhead{$\alpha_{2000}$} & \colhead{$\delta_{2000}$} &
\colhead{z} & \colhead{$L_{X}$(0.3-3.5 keV)} & \colhead{Mass} & \colhead{$r_{200}$} & \colhead{$\sigma$} &
\colhead{$A_{V}$}\\
&\colhead{} &\colhead{} &\colhead{} &\colhead{($10^{-13} \; {\rm erg \; cm}^{-2} {\rm s}^{-1}$)} &\colhead{($10^{15} M_{\sun}$)} &
\colhead{arcsec} & \colhead{km ${\rm s}^{-1}$} & \colhead{mag}}
\startdata
MS 1358+62 &13:59:50.69 &62:31:05.44 &0.3290 &13.82 &$2.15$ &411 &937 &0.075\\
MS 1621+26 &16:25:38.36 &26:27:42.59 &0.4275 &9.74 &$1.52$ &278 &793 &0.106\\
MS 0016+16 &0:21:09.01 &16:42:54.54 &0.5479 &7.98 &$2.59$ &353 &1234 &0.182\\
MS 1054-03 &10:56:59.94 &-3:37:36.48 &0.832 &2.62 &$3.50$ &239 &1170 &0.117\\
\enddata
\tablerefs{
column (4) \citet[]{CNOC1}, and \citet[]{vanD00} for MS1054; column (5) \citet[]{emss}; column (7) \citet[]{cye97} and \citet[]{tran} for MS1054;
columns (6 and 8) \citet[]{mass}, and \citet[]{tran} for MS1054; column (9) NASA/IPAC Extragalactic Database (NED)}
\end{deluxetable}

\clearpage

\begin{deluxetable}{lccccclllll}
\tabletypesize{\tiny}
\tablecaption{Catalog of MS 0015.9+1609 \label{tab0016}}
\tablehead{
\colhead{Name} & \colhead{${\rm {\alpha}_{J2000}}$} & \colhead{${\rm {\delta}_{J2000}}$} &
\colhead{Mag} & \colhead{$Color$\tablenotemark{1}} & \colhead{$z$} & \colhead{B/T} & \colhead{$\mu_D$ / $\mu_B$} &
\colhead{H / Re} & \colhead{AR$_D$ / AR$_B$} & \colhead{PA$_D$ / PA$_B$} \\
&\colhead{(2)} &\colhead{(3)} &\colhead{(4)} &\colhead{(5)} & &\colhead{(6)} &\colhead{(7)} &\colhead{(8)}
&\colhead{(9)} &\colhead{(10)} }
\startdata
A\_3\_2221\_0816 & 00 18 20.9 & 16 24 52.2 & 21.054 & 3.392 & \nodata & 0.08 & 24.610 & $ 2.8 \pm 2.7 $ & $ 0.49 \pm 0.45 $ & 8.9 \\
& & & & & & & 15.108 & $ 0.30 \pm 0.09 $ & $ 0.37 \pm 0.08 $ & 44.8 \\
A\_4\_5402\_5249 & 00 18 21.0 & 16 26 12.9 & 21.594 & 0.924 & 0.385 & 0. & 21.238 & $ 0.55 \pm 0.03 $ & $ 0.38 \pm 0.02 $ & 323.0 \\
B\_2\_5983\_7213$\ast$ & 00 18 33.7 & 16 21 7.9 & 21.949 & 2.753 & \nodata & 1. & 15.953 & $ 0.95 \pm 0.06 $ & $ 0.42 \pm 0.02 $ & 108.9 \\
B\_4\_6632\_6691$\ast$ & 00 18 32.3 & 16 24 1.3 & 21.034 & 1.403 & \nodata & 0.27 & 20.741 & $ 0.32 \pm 0.02 $ & $ 0.89 \pm 0.04 $ & 328.5 \\
& & & & & & & 16.595 & $ 1.21 \pm 0.19 $ & $ 0.29 \pm 0.04 $ & 92.2 \\
C\_2\_0562\_0892$\ast$ & 00 18 45.3 & 16 27 7.3 & 21.870 & 0.671 & \nodata & 0.20 & 21.389 & $ 0.39 \pm 0.02 $ & $ 0.55 \pm 0.03 $ & 91.1 \\
& & & & & & & 14.411 & $ 0.28 \pm 0.05 $ & $ 0.25 \pm 0.05 $ & 3.4 \\
C\_3\_1549\_4203 & 00 18 43.0 & 16 27 24.1 & 20.216 & 2.123 & 0.554 & 0.56 & 19.777 & $ 0.48 \pm 0.02 $ & $ 0.20 \pm 0.01 $ & 355.0 \\
& & & & & & & 12.199 & $ 0.21 \pm 0.01 $ & $ 0.74 \pm 0.04 $ & 2.2 \\
D\_3\_1695\_7261$\ast$ & 00 18 59.2 & 16 25 39.9 & 21.888 & 1.906 & \nodata & 0. & 23.572 & $ 1.15 \pm 0.06 $ & $ 0.57 \pm 0.03 $ & 3.7 \\
D\_4\_1038\_3508 & 00 18 54.5 & 16 25 3.5 & 20.296 & 0.513 & \nodata & 0.26 & 19.202 & $ 0.44 \pm 0.02 $ & $ 0.22 \pm 0.01 $ & 330.2 \\
& & & & & & & 21.617 & $ 20.0 \pm 8.4 $ & $ 0.21 \pm 0.02 $ & 158.1 \\
E\_1\_5836\_6117 & 00 18 36.0 & 16 25 59.1 & 20.800 & 1.907 & 0.554 & 1. & 14.876 & $ 0.44 \pm 0.02 $ & $ 0.93 \pm 0.05 $ & 76.2 \\
E\_3\_0524\_6061 & 00 18 30.2 & 16 26 8.6 & 20.985 & 1.802 & 0.544 & 0.77 & 20.313 & $ 0.21 \pm 0.01 $ & $ 0.46 \pm 0.02 $ & 356.1 \\
& & & & & & & 12.558 & $ 0.22 \pm 0.01 $ & $ 0.66 \pm 0.03 $ & 222.8 \\
\enddata
\tablenotetext{1}{Color is (F814W-F555W)}
\tablecomments{The complete version of this table is in the electronic edition of
the Journal.  The printed edition contains only a sample.}
\end{deluxetable}

\begin{deluxetable}{lcccclllll}
\tabletypesize{\tiny}
\tablecaption{Catalog of MS 1054.4-0321 \label{tab1054}}
\tablehead{
\colhead{Name} & \colhead{${\rm {\alpha}_{J2000}}$} & \colhead{${\rm {\delta}_{J2000}}$} &
\colhead{Mag} & \colhead{$Color$\tablenotemark{1}} &  \colhead{B/T} & \colhead{$\mu_D$ / $\mu_B$} &
\colhead{H / Re} & \colhead{AR$_D$ / AR$_B$} & \colhead{PA$_D$ / PA$_B$} \\
&\colhead{(2)} &\colhead{(3)} &\colhead{(4)} &\colhead{(5)}  &\colhead{(6)} &\colhead{(7)} &\colhead{(8)}
&\colhead{(9)} &\colhead{(10)} }
\startdata
A\_2\_5257\_2843 & 10 56 48.4 & -03 37 31.1 & 20.282 & 0.598 & 0.02 & 24.019 & $ 2.6 \pm 2.0 $ & $ 0.75 \pm 0.39 $ & 200.4 \\
& & & & & &12.274 & $ 0.05 \pm 0.03 $ & $ 0.44 \pm 0.22 $ & 122.1 \\
A\_4\_1255\_5906$\ast$ & 10 56 53.7 & -03 36 35.6 & 22.723 & 1.318 & 0. & 20.26 & $ 0.16 \pm 0.01 $ & $ 0.61 \pm 0.03 $ & 345.2 \\
B\_2\_7326\_1845 & 10 56 53.6 & -03 38 58.9 & 22.086 & 1.318 & 0.22 & 20.235 & $ 0.23 \pm 0.01 $ & $ 0.43 \pm 0.02 $ & 85.5 \\
& & & & & &17.405 & $ 0.82 \pm 0.18 $ & $ 0.41 \pm 0.11 $ & 16.6 \\
B\_4\_4364\_4854 & 10 56 57.5 & -03 37 12.3 & 21.277 & 0.447 & 0. & 21.142 & $ 0.54 \pm 0.03 $ & $ 0.49 \pm 0.02 $ & 304.3 \\
C\_3\_5770\_5963$\ast$ & 10 56 55.7 & -03 38 5.7 & 21.423 & 1.631 & 1. & 15.224 & $ 0.64 \pm 0.03 $ & $ 0.76 \pm 0.04 $ & 179.0 \\
D\_4\_2364\_4815$\ast$ & 10 57 11.7 & -03 37 30.6 & 22.312 & 0.962 & 0.18 & 21.315 & $ 0.27 \pm 0.01 $ & $ 0.73 \pm 0.04 $ & 7.7 \\
& & & & & &14.989 & $ 0.16 \pm 0.09 $ & $ 0.78 \pm 0.21 $ & 297.4 \\
E\_2\_1561\_2215 & 10 57 2.7 & -03 36 50.6 & 21.956 & 1.665 & 1. & 14.445 & $ 0.34 \pm 0.02 $ & $ 0.83 \pm 0.04 $ & 97.4 \\
E\_3\_4427\_6160 & 10 57 1.3 & -03 35 43.6 & 21.234 & 0.411 & 0.11 & 21.013 & $ 0.98 \pm 0.05 $ & $ 0.12 \pm 0.01 $ & 36.8 \\
& & & & & &16.265 & $ 0.44 \pm 0.13 $ & $ 0.58 \pm 0.10 $ & 24.1 \\
F\_2\_5849\_3682$\ast$ & 10 56 57.3 & -03 36 23.9 & 21.177 & 1.011 & 0.01 & 21.334 & $ 0.51 \pm 0.03 $ & $ 0.71 \pm 0.04 $ & 12.2 \\
& & & & & &12.001 & $ 0.02 \pm 0.03 $ & $ 0.5 \pm 1.3 $ & 273.0 \\
F\_4\_2673\_1801$\ast$ & 10 57 0.4 & -03 35 8.3 & 22.295 & 0.706 & 0.11 & 24.273 & $ 2.7 \pm 6.2 $ & $ 0.12 \pm 0.25 $ & 335.3 \\
& & & & & &16.514 & $ 0.43 \pm 0.23 $ & $ 0.26 \pm 0.06 $ & 174.4 \\
\enddata
\tablenotetext{1}{Color is (F814W-F606W)}
\tablecomments{The complete version of this table is in the electronic edition of
the Journal.  The printed edition contains only a sample.}
\end{deluxetable}

\begin{deluxetable}{lccccclllll}
\tabletypesize{\tiny}
\tablecaption{Catalog of MS 1358.4+6245 \label{tab1358}}
\tablehead{
\colhead{Name} & \colhead{${\rm {\alpha}_{J2000}}$} & \colhead{${\rm {\delta}_{J2000}}$} &
\colhead{Mag} & \colhead{$Color$\tablenotemark{1}} & \colhead{$z$} & \colhead{B/T} & \colhead{$\mu_D$ / $\mu_B$} &
\colhead{H / Re} & \colhead{AR$_D$ / AR$_B$} & \colhead{PA$_D$ / PA$_B$} \\
&\colhead{(2)} &\colhead{(3)} &\colhead{(4)} &\colhead{(5)} & &\colhead{(6)} &\colhead{(7)} &\colhead{(8)}
&\colhead{(9)} &\colhead{(10)} }
\startdata
A\_2\_5725\_2798 & 13 59 48.3 & 62 31 57.8 & 20.511 & 0.839 & 0.322 & 0.75 & 20.192 & $ 0.34 \pm 0.02 $ & $ 0.24 \pm 0.01 $ & 176.3 \\
& & & & & & & 12.465 & $ 0.33 \pm 0.02 $ & $ 0.40 \pm 0.02 $ & 174.3 \\
A\_4\_0726\_5554 & 13 59 46.3 & 62 30 25.9 & 20.172 & 0.842 & 0.334 & 1. & 12.166 & $ 0.31 \pm 0.02 $ & $ 0.61 \pm 0.03 $ & 15.3 \\
B\_2\_4601\_2187$\ast$ & 13 59 34.4 & 62 31 47.1 & 21.977 & 0.633 & \nodata & 0. & 20.353 & $ 0.23 \pm 0.01 $ & $ 0.70 \pm 0.03 $ & 332.8 \\
D\_4\_5092\_2352 & 13 59 32.3 & 62 27 42.4 & 17.664 & 0.179 & 0.328 & 0.23 & 25.007 & $ 29.9 \pm 6.7 $ & $ 0.12 \pm 0.28 $ & 107.5 \\
& & & & & & & 11.552 & $ 0.34 \pm 0.02 $ & $ 0.65 \pm 0.03 $ & 245.8 \\
F\_2\_4360\_5659 & 13 59 59.8 & 62 29 33.7 & 19.203 & 0.886 & 0.337 & 0.61 & 20.473 & $ 0.50 \pm 0.02 $ & $ 0.80 \pm 0.04 $ & 85.1 \\
& & & & & & & 12.816 & $ 0.46 \pm 0.02 $ & $ 0.75 \pm 0.04 $ & 80.0 \\
G\_3\_7214\_0725$\ast$ & 14 00 19.9 & 62 29 16.1 & 20.875 & 1.317 & 0.532 & 0.89 & 20.762 & $ 0.22 \pm 0.01 $ & $ 0.33 \pm 0.03 $ & 108.6 \\
& & & & & & & 13.556 & $ 0.47 \pm 0.02 $ & $ 0.44 \pm 0.02 $ & 105.4 \\
I\_3\_7580\_4164 & 14 00 9.7 & 62 34 4.2 & 22.487 & 0.577 & \nodata & 0. & 22.383 & $ 0.39 \pm 0.02 $ & $ 0.96 \pm 0.05 $ & 87.1 \\
J\_2\_7234\_6432$\ast$ & 14 00 8.2 & 62 25 51.8 & 21.527 & 2.396 & \nodata & 0. & 19.627 & $ 0.32 \pm 0.02 $ & $ 0.28 \pm 0.01 $ & 80.5 \\
L\_2\_4827\_1935$\ast$ & 13 59 41.2 & 62 34 20.8 & 21.094 & 0.456 & 0.382 & 0.14 & 20.890 & $ 0.41 \pm 0.02 $ & $ 0.68 \pm 0.03 $ & 75.8 \\
& & & & & & & 19.075 & $ 2.34 \pm 0.19 $ & $ 0.37 \pm 0.03 $ & 100.1 \\
M\_3\_5846\_7043 & 13 59 41.6 & 62 33 30.4 & 20.699 & 0.870 & \nodata & 0.63 & 23.288 & $ 1.51 \pm 0.23 $ & $ 0.28 \pm 0.04 $ & 222.5 \\
& & & & & & & 15.682 & $ 1.13 \pm 0.08 $ & $ 0.45 \pm 0.02 $ & 178.1 \\
\enddata
\tablenotetext{1}{Color is (F814W-F606W) except for galaxies with names starting with J or K, in which case it is (F814W-F450W)}
\tablecomments{The complete version of this table is in the electronic edition of
the Journal.  The printed edition contains only a sample.}
\end{deluxetable}

\begin{deluxetable}{lccccclllll}
\tabletypesize{\tiny}
\tablecaption{Catalog of MS 1621.5+2640 \label{tab1621}}
\tablehead{
\colhead{Name} & \colhead{${\rm {\alpha}_{J2000}}$} & \colhead{${\rm {\delta}_{J2000}}$} &
\colhead{Mag} & \colhead{$Color$\tablenotemark{1}} & \colhead{$z$} & \colhead{B/T} & \colhead{$\mu_D$ / $\mu_B$} &
\colhead{H / Re} & \colhead{AR$_D$ / AR$_B$} & \colhead{PA$_D$ / PA$_B$} \\
&\colhead{(2)} &\colhead{(3)} &\colhead{(4)} &\colhead{(5)} & &\colhead{(6)} &\colhead{(7)} &\colhead{(8)}
&\colhead{(9)} &\colhead{(10)} }
\startdata
A\_1\_3143\_6511 & 16 23 39.7 & 26 35 2.8 & 21.569 & 1.609 & \nodata & 0.48 & 20.581 & $ 0.27 \pm 0.01 $ & $ 0.20 \pm 0.01 $ & 339.7 \\
& & & & & & & 13.672 & $ 0.13 \pm 0.01 $ & $ 0.81 \pm 0.04 $ & 124.7 \\
A\_3\_2629\_2271 & 16 23 35.5 & 26 35 9.6 & 21.635 & 1.510 & \nodata & 0. & 22.906 & $ 0.97 \pm 0.06 $ & $ 0.54 \pm 0.03 $ & 60.2 \\
B\_2\_0999\_0800 & 16 23 40.8 & 26 31 35.1 & 21.432 & 1.774 & \nodata & 0.60 & 20.605 & $ 0.19 \pm 0.01 $ & $ 0.85 \pm 0.04 $ & 74.4 \\
& & & & & & & 15.181 & $ 0.47 \pm 0.03 $ & $ 0.81 \pm 0.06 $ & 8.1 \\
B\_3\_0615\_2903$\ast$ & 16 23 39.4 & 26 31 57.5 & 21.307 & 1.526 & \nodata & 1. & 12.868 & $ 0.37 \pm 0.02 $ & $ 0.30 \pm 0.01 $ & 118.0 \\
C\_2\_1460\_4198 & 16 23 24.9 & 26 37 38.0 & 19.062 & 1.384 & 0.431 & 0.95 & 23.377 & $ 0.80 \pm 0.04 $ & $ 0.72 \pm 0.04 $ & 8.1 \\
& & & & & & & 16.154 & $ 2.66 \pm 0.13 $ & $ 0.87 \pm 0.04 $ & 173.2 \\
C\_4\_1317\_2943 & 16 23 27.3 & 26 38 36.0 & 20.646 & 0.750 & \nodata & 1. & 19.386 & $ 6.56 \pm 0.40 $ & $ 0.69 \pm 0.03 $ & 3.8 \\
D\_2\_2851\_6055$\ast$ & 16 23 48.9 & 26 27 50.4 & 21.859 & 1.518 & \nodata & 0.25 & 21.715 & $ 0.40 \pm 0.02 $ & $ 0.67 \pm 0.03 $ & 92.2 \\
& & & & & & & 13.063 & $ 0.11 \pm 0.02 $ & $ 0.56 \pm 0.09 $ & 45.3 \\
D\_3\_0732\_3391$\ast$ & 16 23 48.9 & 26 29 5.4 & 21.713 & 1.052 & \nodata & 0.33 & 20.519 & $ 0.29 \pm 0.01 $ & $ 0.43 \pm 0.02 $ & 29.0 \\
& & & & & & & 16.167 & $ 1.12 \pm 0.10 $ & $ 0.15 \pm 0.02 $ & 36.4 \\
E\_2\_7010\_6640 & 16 23 35.9 & 26 43 40.5 & 20.393 & 2.329 & \nodata & 0. & 18.798 & $ 0.32 \pm 0.02 $ & $ 0.36 \pm 0.02 $ & 105.2 \\
E\_3\_3666\_2448$\ast$ & 16 23 33.0 & 26 45 1.1 & 19.888 & 1.216 & 0.393 & 0.14 & 20.218 & $ 0.54 \pm 0.03 $ & $ 0.63 \pm 0.03 $ & 95.0 \\
& & & & & & & 11.023 & $ 0.12 \pm 0.01 $ & $ 0.25 \pm 0.01 $ & 29.0 \\

\enddata
\tablenotetext{1}{Color is (F814W-F555W)}
\tablecomments{The complete version of this table is in the electronic edition of
the Journal.  The printed edition contains only a sample.}
\end{deluxetable}

\begin{deluxetable}{lccc}
\tablecaption{Normalization factor \label{tabnorm}}
\tablehead{
\colhead{Cluster} & \colhead{$f_{m}$\tablenotemark{1}} & \colhead{$f_{g}$\tablenotemark{2}} }
\startdata
MS 1358+62 &0.657 &0.477\\
MS 1621+26 &0.314 &0.161\\
MS 0016+16 &0.312 &0.273\\
MS 1054-03 &0.846 &1.00\\
\enddata
\tablenotetext{1}{Fractions of the cluster mass observed, using the density profile of \citet[]{carlberg97}} 
\tablenotetext{2}{Fractions of the cluster galaxies observed, normalized to MS1054.  Factors calculated with the cluster masses from \citet[]{mass} and \citet[]{tran} for MS1054.  These are the fraction of the cluster observed weighted by the mass of the clusters. }
\end{deluxetable}

\begin{deluxetable}{lccccccc}
\tabletypesize{\scriptsize}
\tablecaption{Results of Evolution Models \label{tablum}}
\tablehead{
\colhead{Cluster} & \colhead{N\tablenotemark{1}} & \colhead{$\Delta h$} & 
\colhead{$\Delta \mu_{0}\tablenotemark{2}$} & \colhead{$\Delta \mu_{0}(95\%)\tablenotemark{3}$}
& \colhead{med$<h>$ \tablenotemark{4}} & 
\colhead{$\sigma_h$} & \colhead{$\Delta \mu_{0}$ (env)\tablenotemark{5}}}
\startdata
MS 1358+62 &15 &0.55 &1.50 & 0.70---1.76 & 1.40 &0.41 &0.8-1.4\\
MS 1621+26 &5 &0.37 &0.43  & 0.0---0.87  & 1.14 &0.54 &0.7-1.2\\
MS 0016+16 &9 &0.33 &0.41 & 0.0---0.80   & 1.53 &0.30 &0.1-0.7\\
MS 1054-03 &33 &\nodata & \nodata &\nodata &1.29 &0.38 &\nodata\\
\enddata
\tablenotetext{1}{``Equivalent'' number of small bright disks used for each cluster.  They were calculated with the normalization factors and based on the count of 33 small bright disks in MS1054}
\tablenotetext{2}{Shift required in the surface brightness as determined by the counts normalization 
technique and its 95\% confidence interval.}
\tablenotetext{3}{The 95\% confidence interval in the surface brightness shift.} 
\tablenotetext{4}{Median of the size distribution of the small bright disks after the shift in surface brightness, also the standard deviation of that distribution}
\tablenotetext{5}{Range for the shift in the surface brightness evaluated with the upper envelope technique.}
\end{deluxetable}


\begin{thebibliography}{}
\bibitem[Balogh, Navarro \& Morris(2000)]{balogh00} Balogh, M.L., Navarro, J.F., \& Morris, S.L.\ 2000, \apj, 540, 113
\bibitem[Balogh et al.(1999)]{balogh99} Balogh, M.L., Morris, S.L., Yee, H.K.C., Carlberg, R.G., \& Ellingson, E.\ 1999, \apj, 527, 54
\bibitem[Balogh et al.(1998)]{balogh98} Balogh, M.L., Schade, D., Morris, S.L., Yee, H.K.C., Carlberg, R.G., \& Ellingson, E.\ 1998, \apjl, 504, L75
\bibitem[Balogh et al.(1997)]{balogh97} Balogh, M.L., Morris, S.L., Yee, H.K.C., Carlberg, R.G., \& Ellingson, E.\ 1997, \apjl, 488, L75
\bibitem[Bertin \& Arnouts(1996)]{bert96} Bertin, E. \& Arnouts, S. 1996, \aap, 117, 393
\bibitem[Bower, Lucey, \& Ellis (1992)]{bower92}Bower, R.,Lucey, J., \& Ellis, R. 1992 \mnras, 254, 601
D., Davies, R. L., Faber, S. M., Terlevich, R. J., \& Wegner, G. 1987, \apj, 313, 42
\bibitem[Butcher \& Oemler(1984)]{bo84} Butcher, H. \& Oemler, A., Jr. 1984, \apj, 285, 426 
\bibitem[Carlberg et al.(1999)]{carlberg1}Carlberg, R., Yee, H., Morris, S., Lin, H., Ellingson, E., Patton, D., Sawicki, M., \& Shepherd, C. 1999, \apj, 516, 552
\bibitem[Carlberg et al.(1997a)]{carlberg97a} Carlberg, R. G., Morris S. L., Yee, H. K. C., \& Ellingson, E. 1997, \apjl, 479, 19
\bibitem[Carlberg et al.(1997b)]{carlberg97} Carlberg, R. G., Yee, H. K. C., Ellingson, E., Morris, S., Abraham, R., Grabel, P., Pritchet, C. J., Smecker-Hane, T., Hartwick, F. D. A., Hesser, J. E., Hutchings, J. B., \& Oke, J. B. 1997, \apjl, 485, L13 
\bibitem[Carlberg, Yee, \& Ellingson(1997)]{cye97} Carlberg, R. G., Yee, H. K. C., \& Ellingson, E. 1997, \apj, 478, 462
\bibitem[Carlberg et al.(1996)]{mass} Carlberg, R. G., Yee, H. K. C., Ellingson, E., Abraham, R. G., Gravel, P., Morris, S., \& Pritchet, C. J. 1996, \apj, 462, 32
\bibitem[Coleman, Wu, \& Weedman (1980)]{coleman} Coleman, G. D., Wu C.-C., Weedman, D. W. 1980, \apjs, 43, 393
\bibitem[Couch, Ellis, Sharples, \& Smail (1994)]{couch94}Couch, W., Ellis, R., Sharples, R., \& Smail, I. 1994, \apj, 430, 121
\bibitem[Crampton et al.(2002)]{crampton02} Crampton, D., Schade, 
D., Hammer, F., Matzkin, A., Lilly, S.~J., \& Le F{\` e}vre, O.\ 2002, 
\apj, 570, 86 
\bibitem[Dalcanton, Spergel, \& Summers(1997)]{dalcanton} Dalcanton, J. J., 
Spergel, D. N., \& Summers, F. J. 1997, \apj, 482, 659
\bibitem[Djorgovski \& Davis (1987)]{djorg87}Djorgovski, S., \& Davis, M. 1987 \apj, 313, 59
\bibitem[Donahue et al.(1998)]{donahue} Donahue, M., Voit, G. M., Gioia, I.,
Luppino, G., Hughes, J. P., \& Stocke, J. T. 1998, \apj, 502, 550
\bibitem[Dominguez, Muriel, \& Lambas (2001)]{dominguez} Dominguez, M., Muriel, Hernan, \& Lambas D. G. 2001, \apj, 121, 1266
\bibitem[Malle\'n-Ornelas et al (1999)]{mallen99}Mallen-Ornelas, G., Lilly, S., Crampton, D.,
\& Schade, D. 1999 \apj, 518, 83
\bibitem[Dressler(1980)]{dress80} Dressler, A. 1980, \apj, 236, 351
\bibitem[Dressler(1984)]{dress84} Dressler, A. 1984, \araa, 22, 185
\bibitem[Dressler et al. (1987)]{dress87}Dressler, A., Lynden-Bell, D., Burstein,
D., Davies, R. L., Faber, S. M., Terlevich, R. J., \& Wegner, G. 1987, \apj, 313, 42
\bibitem[Dressler et al (1994)]{dress94}Dressler, A., Oemler, A., Butcher, H. \& Gunn, J. 1994 \apj, 430, 107
\bibitem[Dressler et al.(1997)]{dress97} Dressler, A., Oemler, A. Jr., Couch, W. J.,
Smail, I., Ellis, R. S., Barger, A., Butcher, H., Poggianti, B. M., \& 
Sharples, R. M. 1997, \apj, 490, 577
\bibitem[Ellingson et al.(1997)]{CNOC3} Ellingson, E., Yee, H. K. C., Abraham, R. G.,
Morris, S. L., Carlberg, R. G., \& Smecker-Hane, T. A. 1997, \apjs, 113, 1
\bibitem[Ellingson et al.(1998)]{CNOC6} Ellingson, E., Yee, H. K. C., Abraham, R. G.,
Morris, S. L., \& Carlberg, R. G. 1998, \apjs, 116, 247
\bibitem[Ellingson et al.(2001)]{ellingson01} Ellingson, E., Lin, H., Yee, H. K. C., \& Carlberg, R. G.
2001, \apj, 547, 609
\bibitem[Ellis \& Jones(2002)]{ellis02} Ellis, S. C., \& Jones, L. R. 2002, \mnras, 330, 631
\bibitem[Ellis et al.(1997)]{ellis97} Ellis, R. S., Smail, I., Dressler, A., Couch, W. J., 
Oemler, A. Jr., Butcher, H., \& Sharples, R. M. 1997, \apj, 483, 582 
\bibitem[Fisher et al.(1998)]{Fish98} Fisher, D., Fabricant, D., Franx, M., \& 
van Dokkum, P. G. 1998, \apj, 498, 195
\bibitem[Forbes et al.(1996)]{forbes96} Forbes, D. A., Phillips, A. C., Koo, D. C., \& 
Illingworth, G. D. 1996, \apj, 462, 89
\bibitem[Gehrels (1986)]{gehrels} Gehrels, N. 1986 \apj, 303, 336
\bibitem[Gioia et al.(1990)]{emss} Gioia, I. M., Maccacaro, T., Schild, R. E., Wolter, A.,
Stocke, J. T., Morris, S. L., \& Henry, J. P. 1990, \apjs, 72, 567
\bibitem[Gladders, Yee, \& Ellingson(2002)]{gladders02} Gladders, 
M.~D., Yee, H.~K.~C., \& Ellingson, E.\ 2002, \aj, 123, 1 
\bibitem[Guzma\'n et al (1996)]{guzman96} Guzman, R., Koo, D., Faber, S., Illingworth, G.,
Takamiya, M., Kron, R., \& Bershady, M. 1996 \apj, 460, 5
\bibitem[Hoekstra, Franx, \& Kuijken (2000)]{hoekstra} Hoekstra, H., Franx, M., \& Kuijken, K. 2000, \apj, 532, 88
\bibitem[Hubble \& Humason(1931)]{hubble31} Hubble, E. \& Humason, M. L. 1931, \apj, 74, 43
\bibitem[Jeltema et al.(2001)]{jeltema} Jeltema, T. E., Canizares, C. R., Bautz, M. W., Malm, M. R., Donahue, M., Garmire, G. P.  2001, \apj, 562, 124 
\bibitem[Kelson et al.(1997)]{kelson} Kelson, D. D., van Dokkum, P. g., Franx, M., Illingworth, G. D., \& Fabricant, D. 1997, \apjl, 478, L13
\bibitem[Lavery \& McClure (1992)]{lavery92}Lavery, R. Pierce, M., \& McClure, R. 1992 \aj, 104, 2067
\bibitem[Lavery \& Henry (1994)]{lavery94}Lavery, R. \& Henry, J. 1994 \apjl, 426, L524
\bibitem[Lilly et al.(1995)]{cfrs} Lilly, S. J., Le Fevre, O., Crampton, D., Hammer, F., \& Tresse, L. 1995, \apj, 455, 50
\bibitem[Oemler(1974)]{oemler74} Oemler, A. Jr. 1974, \apj, 194, 1 
\bibitem[Oemler, Dressler, \& Butcher (1997)]{oemler97}Oemler, A., Dressler, A., \& Butcher, H. 1997 \apj, 474, 561
\bibitem[Mao, Mo, \& White(1998)]{MoWhite} Mao, S., Mo, H.~J., 
\& White, S.~D.~M.\ 1998, \mnras, 297, L71 
\bibitem[Marleau \& Simard(1998)]{marleau} Marleau, F. R., \& Simard, L. 1998, \apj, 507, 585
\bibitem[Martel, Premade, \& Matzner(1998)]{martel98} Martel, H., Premadi, P., \& 
Matzner, R. 1998, \apj, 497, 512 
\bibitem[Melnick \& Sargent(1977)]{mel77} Melnick, J. \& Sargent, W. L. W.
1977, \apj, 215, 401
\bibitem[Monet et al.(2003)]{USNO} Monet, D.G., et al.\ 
2003, \aj, 125, 984 
\bibitem[Moore et al (1996)]{moore96} Moore, B., Katz, N., Lake, G., Dressler, A., Oemler, A. Jr. 1996 Nature, 379, 613
\bibitem[Moore, Lake, \& Katz (1998)]{moore98} Moore, B., Lake, G.,\& Katz, N. 1998 \apj, 495, 139
\bibitem[Navarro, Frenk, \& White(1995)]{navarro} Navarro, J. F., Frenk, C. S., 
\& White, S. D. M. 1995, \mnras, 275, 720
\bibitem[Poggianti et al.(1999)]{poggianti99} Poggianti, B.M., Smail, I., Dressler, A., Couch, W.J., Barger, A.J., Butcher, H., Ellis, R.S., \& Oemler, A.J.\ 1999, \apj, 518, 576
\bibitem[Rakos \& Schombert (1995)]{rakos95}Rakos, J. \& Schombert, J. 1995 \apj, 439, 47
\bibitem[Roche et al.(1998)]{roche} Roche, N., Ratnatunga, K., Griffiths, R. E., Im, M., \& 
Naim, A. 1998, \mnras, 293, 157 
\bibitem[Rosati et al.(1998)]{rosati98}Rosati, P., della Ceca, R., Norman, C., Giacconi, R.
1998 \apjl, 492, L21
\bibitem[Sachs(1984)]{stat84} Sachs, L. 1984, Applied Statistics. Second edition,
Springer-Verlag, New York, 373
\bibitem[Schade et al.(1995)]{Schade95} Schade, D., Lilly, S. J., Crampton, D.,
Hammer, F., Le F\`{e}vre, O., \& Tresse, L. 1995, \apj, 451, L1
\bibitem[Schade et al.(1996)]{Schade96} Schade, D., Lilly, S. J., Le F\`{e}vre, O.,
Hammer, F., \&  Crampton, D. 1996, \apj, 464, 79
\bibitem[Schade et al.(1996b)]{Schade96cnocE} Schade,D., Carlberg, R., Yee, H., Lop\`{e}z-Cruz, O.
\& Ellingson, E. 1996, \apj, 464, 163
\bibitem[Schade et al.(1996c)]{Schade96b} Schade, D., Carlberg, R. G., Yee, H. K. C., \& 
Lop\`{e}z-Cruz, O. 1996, \apjl, 465, L103
\bibitem[Schade, Barrientos, \& Lop\`{e}z-Cruz(1997)]{Schade97} Schade, D., Barrientos, L. F., 
\& Lop\`{e}z-Cruz, O. 1997, \apjl, 477, L17
\bibitem[Schade et al.(1999)]{Schade99} Schade, D., et al.\ 
1999, \apj, 525, 31 
\bibitem[Tody (1993)]{iraf} Tody, D. 1993, "IRAF in the Nineties" in {\it Astronomical Data Analysis Software and Systems II}, A.S.P. Conference Ser., Vol 52, eds. R.J. Hanisch, R.J.V. Brissenden, \& J. Barnes, 173 
\bibitem[Tran et al.(1999)]{tran} Tran, K. H., Kelson, D. D., van Dokkum, P., Franx, M., Illingworth, G. D., \& Magee, D. 1999, \apj, 522, 39
\bibitem[van den Bergh(1990)]{vandenbergh} van den Bergh, S. 1990, \apj, 348, 57
\bibitem[van Dokkum et al.(1998a)]{vanD98} van Dokkum, P. G.,
    Franx, M., Kelson, D. D., Illingworth, G. D., Fisher, D., \& 
    Fabricant, D.  1998, \apj, 500, 714
\bibitem[van Dokkum et al.(1998b)]{vanD98b} van Dokkum, P. G., Franx, M., Kelson, D. D., \& Illingworth, G. D. 1998, \apjl, 504, L17
\bibitem[van Dokkum et al.(1999)]{vanD99}van Dokkum, P. G., Franx, M., Fabricant, D.,
Kelson, D. D., Illingworth, G. D. 1999 \apj, 520, L95
\bibitem[van Dokkum et al.(2000)]{vanD00} van Dokkum, P. G., Franx, M., Fabricant, D.,
Illingworth, G. D.,  Kelson, D. D. 2000, \apj, 541, 95
\bibitem[Whitmore, Gilmore, \& Jones(1993)]{whit93} Whitmore, B. C., Gilmore, D. M., 
\& Jones, C. 1993, \apj, 407, 489
\bibitem[Yee, Ellingson \& Carlberg(1996)]{CNOC1} Yee, H. K. C., Ellingson, E., \& 
Carlberg, R. G. 1996 \apjs, 102, 269
\bibitem[Yee et al.(1998)]{CNOC4} Yee, H. K. C., Ellingson, E., Morris, S. L., \& 
Carlberg, R. G. 1998, \apjs, 116, 211

\end{thebibliography}
\end{document}